# Gravitational moduli forces in small nuclei and analytical computation of the Newton constant*

Constantinos G. Vayenas**, Stamatia Giannareli and Stamatios Souentie

*LCEP, Caratheodory 1, St.,*

*University of Patras,*

*Patras GR 26504, Greece*

**Abstract**

The magnitude of gravity at subatomic distances was investigated in the nuclear size range by examining the vibrational kinetic energy of baryons and quarks, treated as waveparticles and strings, in the $^4$He and $^2$H nuclei. Due to the relativistic increase in mass and concomitant wave particle confinement at high vibrational energies, it is found that gravitational moduli forces are significant and comparable in size with the strong interaction forces exerted between color quarks. This leads to analytical expressions for the gravitational constant and for the binding energies of the $^4$He and $^2$H nuclei in close agreement with experiment.

---

* Based on a preprint under consideration for publication at the Intl. Journal of Modern Physics A

** E-mail: cat@chemeng.upatras.gr

## 1. *Introduction*

The computation of the gravitational constant, G, in terms of the fundamental physical constants (h, c, $\varepsilon_o$, e, $m_o$), where e and $m_o$ are the unit charge e and the unit rest mass, has been a long sought goal.[1-3] Current research in this area, commonly known as quantum gravity,[4] focuses primarily on general relativity and string theory.[4-6]

There is increasing evidence and discussion, particularly in the context of brane theory,[6-8] that gravitational forces may be much stronger than anticipated from Newton's law at short, submillimeter distances[7-13] and that there exist extra spatial dimensions curled up into small spaces leading to the formation of "moduli" fields which generate gluon moduli forces similar to but stronger than Newtonian gravity.[7,8] It is also thought that relativity causing a warping of space around branes may lead to very strong moduli forces[10-12] without causing macroscopically significant deviations from Newton's $R^{-2}$ law.[10-12,14] There is also evidence that gravity may be able to cause quantum effects with ultracold neutrons.[15,16]

The combined potential energy, V, due to a modulus force and Newtonian gravity is usually[14] expressed as:

$$V = -\int dr_1 \int dr_2 \frac{G\rho_1(r_1)\rho_2(r_2)}{r_{12}}\left[1+\beta\exp(-r_{12}/\lambda_c)\right] \qquad (1)$$

where the first term expresses Newton's universal gravitational law with G the gravitational constant, $r_{12}$ is the distance between two points $r_1$ and $r_2$ in the test masses and $\rho_1$ and $\rho_2$ are the mass densities of the two bodies. The second term expresses the Yukawa potential with $\beta$ the strength of the new modulus force relative to gravity and $\lambda_c$ is the range. The latter expresses the Compton length, $\lambda_c = h/mc$, corresponding to mass m. Recent experimental[17] and theoretical[14,15] studies have set upper limits for the parameter $\beta$ in terms of $\lambda_c$ and have shown that although new forces can be excluded for $\lambda$ ranges from 200 $\mu m$ to nearly a light-year,[14] limits on new forces become[14] very rapidly poor at distance below 200 $\mu m$.

Based on some of these ideas we have recently[19] explored the existence of gravity-related strong forces in nuclear environments where $\lambda_c$ is in the order of $10^{-15}$ m. We found that when protons and neutrons in small nuclei are treated as harmonic oscillators, then two roots exist for their maximum vibrational velocity, $v_{max}$, and their corresponding vibrational energies and de Broglie wavelengths. One root corresponds to negligible relativistic corrections ($v_{max} = (\alpha / 2\pi)c$, $\gamma \approx 1$, $m = m_o$), while the second root corresponds to very significant relativistic corrections ($v_{max} \approx c$, $\gamma \approx (\alpha / 2\pi)^{-2}$, $m = m_o\gamma$), where $\gamma$ is the time-averaged value of the Lorentz factor $(1 - v^2 / c^2)^{-1/2}$ during each nucleon oscillation and $\alpha$ is the fine structure constant $(= e^2 / \varepsilon c\hbar)$. The first root corresponds[19,20] to energies of proton interactions in chemistry (0.2 to 20 eV). In the case of the second root, which corresponds to energies of ~1 GeV, the pronounced relativistic increase in mass and concomitant pronounced increase in the harmonic oscillator force constant makes, surprisingly, the attractive gravitational forces comparable in magnitude with strong interaction forces and with the repulsive Coulombic forces between protons. This leads to exact analytical expressions for the binding energies, $E_b$, of small nuclei (e.g. $^4$He and $^2$H) and for the gravitational constant:

$$E_{b,{}^4He} = 4\alpha m_o c^2 \left[(1/5) - (4/3)128\alpha\right] = 28.43 \text{ MeV} \tag{2}$$

$$E_{b,{}^2H} = 2\alpha m_o c^2 \left[(2/15) + 4\alpha\right] = 2.225 \text{ MeV} \tag{3}$$

$$G = (2/15)(e^2 / \varepsilon m_o^2)(\alpha / 2\pi)^{12} = 6.672 \cdot 10^{-11} \text{ m}^3 / \text{kgs}^2 \tag{4}$$

All the above three analytical expressions are in quantitative agreement with experiment, i.e. with the currently recommended CODATA values.[21,22]

Interestingly, as discussed in the last section of this work, when examining the above results in terms of equation (1), one obtains $\lambda_c$ values between $10^{-18}$ m (the size of color quarks) and $10^{-24}$ m (the size of a relativistic Planck length) and $\beta \approx 2 \cdot 10^{35}$ and these $(\lambda_c, \beta)$ points fall near to the extension of the average $\beta$ vs $\lambda_c$ line established from the previous theoretical and experimental studies at much longer distances.[14]

In this paper we present an alternative and more rigorous derivation of equations (2) to (4) and show the close relationship between the relativistic gravitational forces in nuclei and the forces between color quarks. We also show a close relationship between ion-induced dipole energies and charges in nuclei and the energies and charges of u and d quarks. In the present derivation, which is based primarily on Heisenberg's uncertainty principle and relativity, the kinetic energy in the Hamiltonians of nucleons and quarks is expressed using the de Broglie equation and accounting for relativistic effects.

## *2. Coulombic and strong interaction energies in a $^4$He nucleus*

### *2.1. Nuclear binding energy*

The total energy a $^4He$ nucleus at rest is $4m_{o,He}c^2$, where $m_{o,He}$ is the average rest mass of protons and neutrons in the nucleus.

Denoting by $m_{o,p}$ and $m_{o,n}$ the rest mass of a proton and a neutron far from the nucleus, one can express the binding energy of the $^4He$ nucleus via:

$$E_b = (2m_{o,p} + 2m_{o,n})c^2 - 4m_{o,He}c^2 \qquad (5)$$

This is shown schematically in Figure 1. The computation of $E_b$ and the actual energy-distance curve shown in Figure 1 is discussed in section 6.

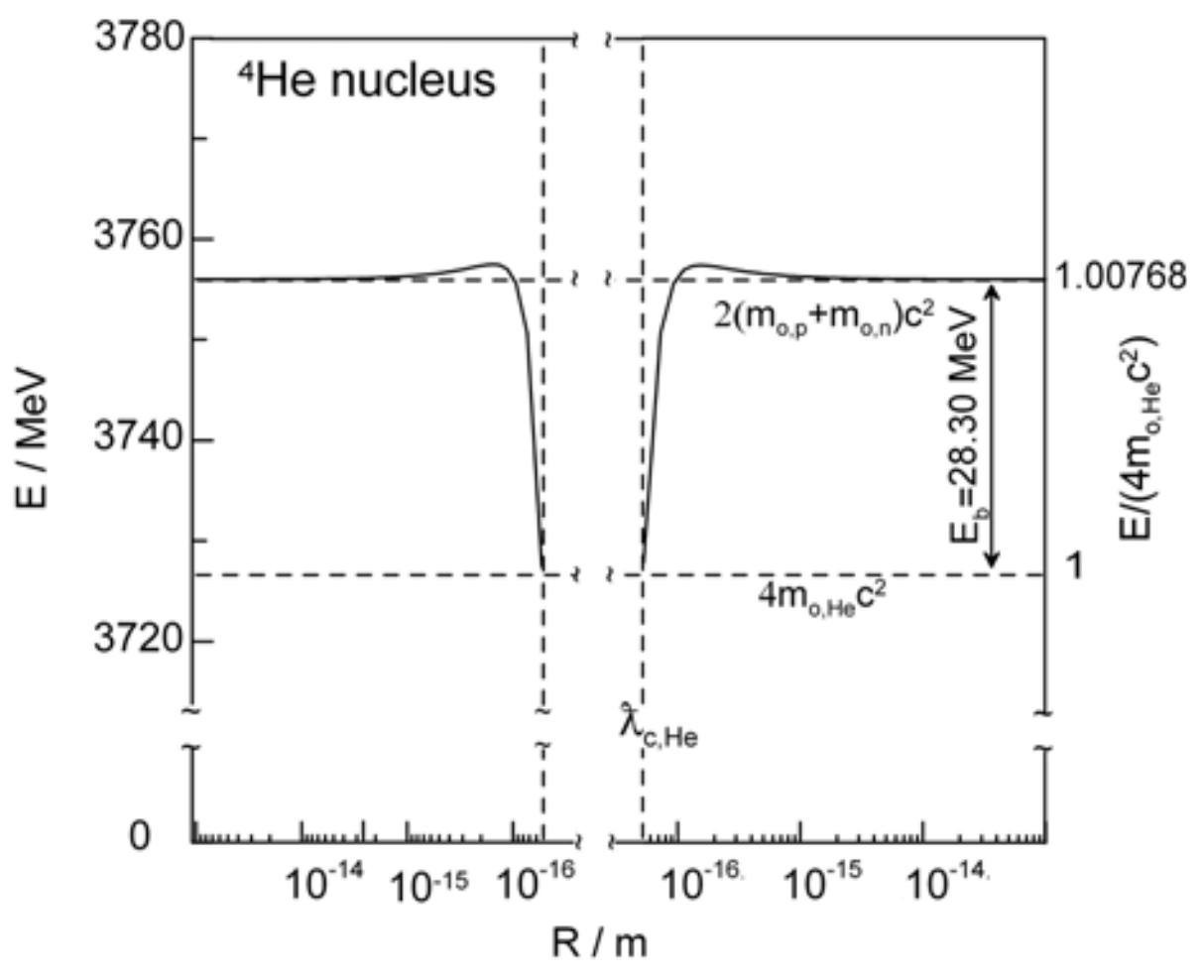

**Fig. 1.** Schematic of the definition of the binding energy of the $^4He$ nucleus; $ƛ_{c,He}$ is the $^4He$ Compton length. The actual E vs R curve shown is discussed in section 6.

### *2.2. Coulombic and attractive interaction energies*

We then consider the total energies, $E_{TOT,p}$ and $E_{TOT,n}$ of a proton and neutron respectively in the $^4He$ nucleus. The proton has a Coulombic energy $\bar{E}_C (> 0)$ due to its interaction with the other proton and three attractive interaction energies, each denoted $\bar{E}_{attr} (< 0)$, due to its strong interaction forces with the other three nucleons. The neutron has only three attractive interaction energies $\bar{E}_{attr}$.

These strong interactions are known to result primarily from interactions of the color quarks u and d of the nucleons. The proton is a uud particle and the neutron is a udd particle.[23,24]

Thus one can write:

$$E_{TOT,p} = m_{o,p,He}c^2 + \bar{E}_C + 3\bar{E}_{attr} + \bar{E}_K = m_{o,He}c^2 \tag{6}$$

$$E_{TOT,n} = m_{o,n,He}c^2 + 3\bar{E}_{attr} + \bar{E}_K = m_{o,He}c^2 \tag{7}$$

where $m_{o,p,He}$ and $m_{o,n,He}$ are the proton and neutron rest masses inside the $^4He$ nucleus, and $\bar{E}_K$ the time averaged kinetic energy of each nucleon due to its vibration around some average equilibrium position as analyzed below. This vibration causes the strong interaction energy to also oscillate in time and $\bar{E}_{attr}$ denotes the time average value of this interaction energy. In equations (6) and (7) we have assumed equal kinetic energies of the nucleons and equal attractive interaction energies, $\bar{E}_{attr}$, for all pairwise nucleon-nucleon interactions.

From equations (6) and (7) it follows:

$$m_{o,n,He}c^2 - m_{o,p,He}c^2 = \bar{E}_C \tag{8}$$

and thus equation (7) can also be written as:

$$E_{TOT,n} = m_{o,p,He}c^2 + \bar{E}_C + 3\bar{E}_{attr} + \bar{E}_K = m_{o,He}c^2 \tag{9}$$

i.e., the neutron behaves as if it had the rest mass of the proton plus a Coulombic energy $\bar{E}_C$. Setting $m_{o,p,He} = m_{o,He}$ in equation (6) or (9) one obtains:

$$\bar{E}_C + 3\bar{E}_{attr} + \bar{E}_K = 0 \tag{10}$$

Anticipating that the potential energy of the proton, $\bar{E}_C + 3\bar{E}_{attr}$, oscillates between $\bar{E}_C$ and $-\bar{E}_C$, it follows that the maximum value of the kinetic energy $E_{K,max}$, i.e. the total energy of the oscillator, is $2\bar{E}_C$, and thus:

$$\bar{E}_K = (1/2)E_{K,max} = \bar{E}_C \tag{11}$$

and also that:

$$-3\overline{E}_{attr} = 2\overline{E}_C \tag{12}$$

Figure 2a shows the values of $3\overline{E}_{attr}$, $\overline{E}_K$, $m_{o,p,He}c^2$ and $m_{o,n,He}c^2$ computed from equations (8) to (12) for an arbitrary value of $\overline{E}_C$. The latter can be computed from:

$$\overline{E}_C = e^2 / \varepsilon R \tag{13}$$

where R is the average distance between the two protons and $\varepsilon = 4\pi\varepsilon_o\varepsilon_r$, with $\varepsilon_r = 1$ for vacuum.

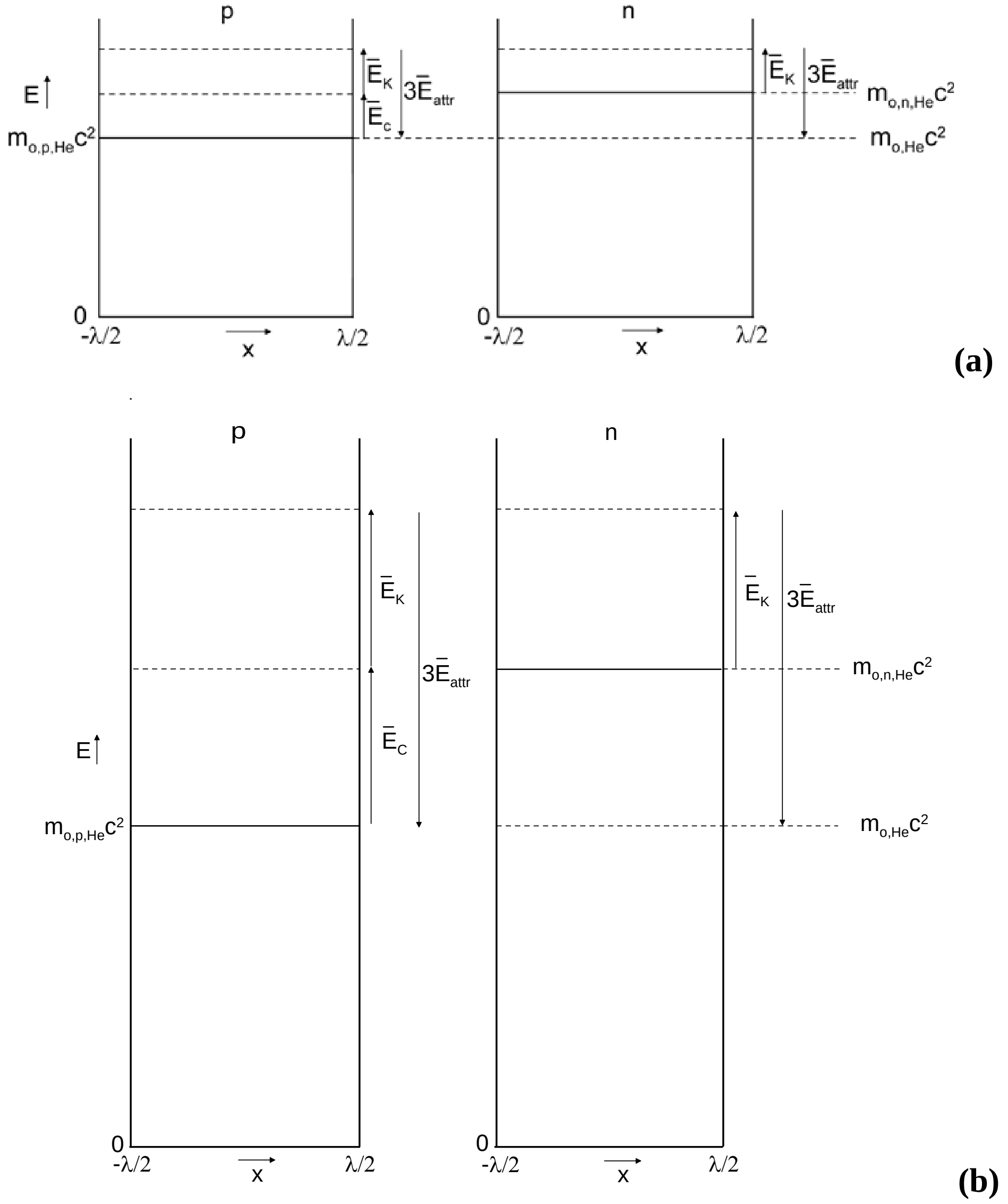

**Fig. 2.** Energy diagrams of a proton (left) and a neutron (right) in a $^4$He nucleus with a time-averaged strong interaction energy $3\overline{E}_{attr}$ due to the attractive interactions with the other three nucleons; $\overline{E}_C$ is the Coulombic repulsion energy and $\overline{E}_K$ is the average kinetic energy of the oscillator. (a) : $2\overline{E}_C \ll m_{o,He}c^2$ ; (b): $2\overline{E}_C = m_{o,He}c^2$.

Since the $^4$He nucleus under consideration is at rest, its total energy is $4m_{o,He}c^2$ and thus the total energy of each oscillating proton or neutron is limited to $m_{o,He}c^2$, which implies that its average kinetic energy, $\bar{E}_K$, is limited to $(1/2)m_{o,He}c^2$, i.e.:

$$E_{K,max} \le m_{o,He}c^2 \quad ; \quad \bar{E}_K \le (1/2)m_{o,He}c^2 \tag{14}$$

Thus at the limit $\bar{E}_K = (1/2)m_{o,He}c^2$, it follows from equations (11) and (12) that:

$$\bar{E}_K = \bar{E}_C = -(3/2)\bar{E}_{attr} = (1/2)m_{o,He}c^2 \quad ; \quad \bar{E}_{attr} = (2/3)\bar{E}_C \tag{15}$$

as shown schematically in figure 2b.

The last equation (15) is strongly reminiscent of the forces exerted between color quarks.[23,24] Thus, for example, the interaction energy between a u quark (charge +(2/3)e) and d quark (charge –(1/3)e) is $-(2/9)\bar{E}_C$. Three such attractive interactions add to the $-(2/3)\bar{E}_C$ of equation (15). And if, as another example, one considers the interaction of a proton with two d quarks of a neighboring neutron, then again the interaction energy is $-(2/3)\bar{E}_C$. This is further discussed in sections 5 and 6.

### *2.3. Attractive interactions and de Broglie wavelength*

Using the dual wave-particle of each nucleon, we write its Hamiltonian by expressing the kinetic energy in terms of the de Broglie equation and corresponding de Broglie wavelength $\lambda$. Thus for the proton and the neutron, respectively, it is:

$$H_p(\lambda, R) = \left( \frac{h^2}{2m\lambda^2}, \frac{(e^2/\varepsilon)}{R} + 3\bar{E}_{attr}(R) \right) \quad ; \quad H_n(\lambda, R) = \left( \frac{h^2}{2m\lambda^2}, 3\bar{E}_{attr}(R) \right) \tag{16}$$

In these expressions the zeroes of the potential energies are at $m_{o,p,He}c^2$ and $m_{o,n,He}c^2$ respectively. If these zeroes are set at $m_{o,p,He}c^2 - \bar{E}_C$ and $m_{o,n,He}c^2 - 2\bar{E}_C$ respectively, then the potential energy term in both equations expresses the potential energy, $E_{P,osc}$, of the corresponding oscillators, i.e.:

$$H_{osc}(\lambda, R) = \left( \frac{h^2}{2m\lambda^2}, \frac{2(e^2/\varepsilon)}{R} + 3\bar{E}_{attr}(R) \right) \tag{17}$$

for both the proton and the neutron. Thus $E_{P,osc}$ oscillates between zero and $2\bar{E}_C$ and in both cases its average value, $\bar{E}_{P,osc}$, is $\bar{E}_C (= \bar{E}_K)$.

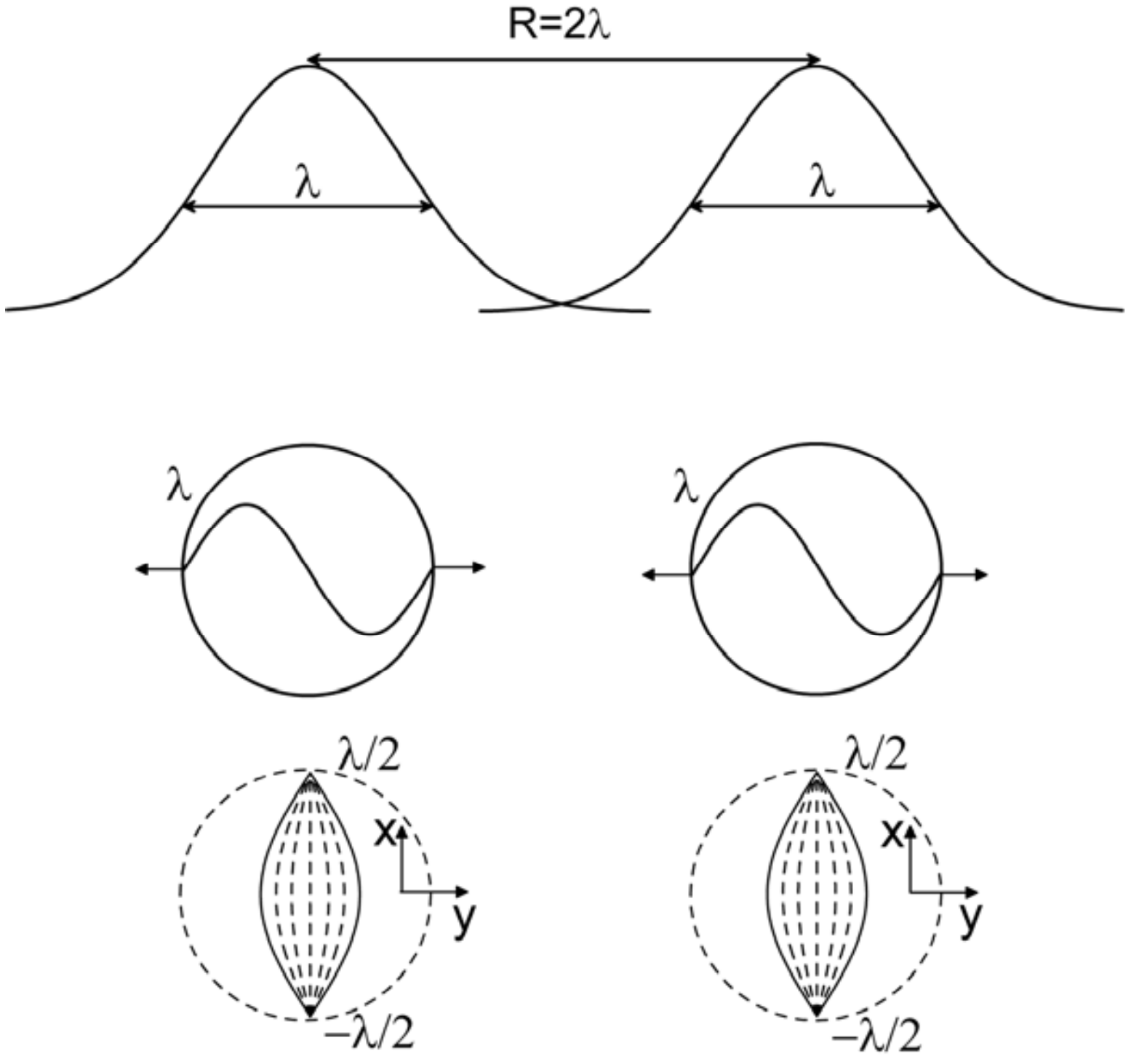

**Fig. 3.** Schematic of two interacting wave particles of de Broglie wavelength $\lambda$ at a distance R=2$\lambda$ viewed as probability distributions (top) as two wavepackets (middle) and as two string oscillators (bottom).

Thus, also relating the average particle distance, R, with the de Broglie wavelength, $\lambda$, via $R = 2\lambda$ (Figure 3) one obtains for both particles:

$$\left\langle H_{osc}(\lambda) \right\rangle = \frac{h^2}{2m\lambda^2} + \frac{e^2}{2\varepsilon\lambda} = E_T \le m_{o,He}c^2 \tag{18}$$

This equation can be used to determine $\lambda$ and $E_T$. Thus, since the average kinetic and potential energy of each nucleon oscillator must be equal, it follows from equation (18) that:

$$\frac{h^2}{2m\lambda^2} = \frac{e^2}{2\varepsilon\lambda} \le (1/2)m_o c^2 \tag{19}$$

From the first equation, also accounting for equation (12), one obtains:

$$\lambda = \frac{\varepsilon h^2}{me^2} \quad ; \quad \bar{E}_C = -\left(\frac{3}{2}\right)\bar{E}_{attr} = \frac{e^2}{2\varepsilon\lambda} = \frac{me^4}{2\varepsilon^2 h^2} \tag{20}$$

Denoting $m_{o,He} = m_o$ and setting $m = m_o$ these equations give the following root, denoted $\lambda_1$:

$$\lambda_1 = \frac{\varepsilon h^2}{m_o e^2}(= (\alpha/2\pi)^{-1}\lambda_C = 1.14\cdot 10^{-12}\ \text{m}) \ ; \ \bar{E}_{C,1} = \frac{m_o e^4}{2\varepsilon^2 h^2}(= \frac{1}{2} m_o c^2 (\alpha/2\pi)^2 = 635\ \text{eV}) \tag{21}$$

where $\lambda_c (= h/m_o c = 1.32\cdot 10^{-15}\ \text{m})$ is the proton or neutron Compton wavelength and $\alpha$ is the fine structure constant $(\alpha = e^2/\varepsilon c\hbar = 1/137.035)$. Clearly this root $(\lambda_1 = 1.14\cdot 10^{-12}\ \text{m},\ E_{C,1} = 635\ \text{eV}\ \text{for}\ \varepsilon_r = 1)$ satisfies the inequality (19). This is the only root above the Compton length $\lambda_c$ (Figure 4). Heisenberg's uncertainty principle dictates that $\Delta p \Delta x > h$. Expressing $\Delta p$ as mc and $\Delta x$ as the de Broglie wavelength $\lambda$, one thus obtains $m\lambda > h/c$ (Figure 4). The line $m\lambda = h/c$, which defines the upper limit of the forbidden region, contains the point $(\lambda_c, m_o)$. Thus for $\lambda < \lambda_c$ it is necessarily $m > m_o$.

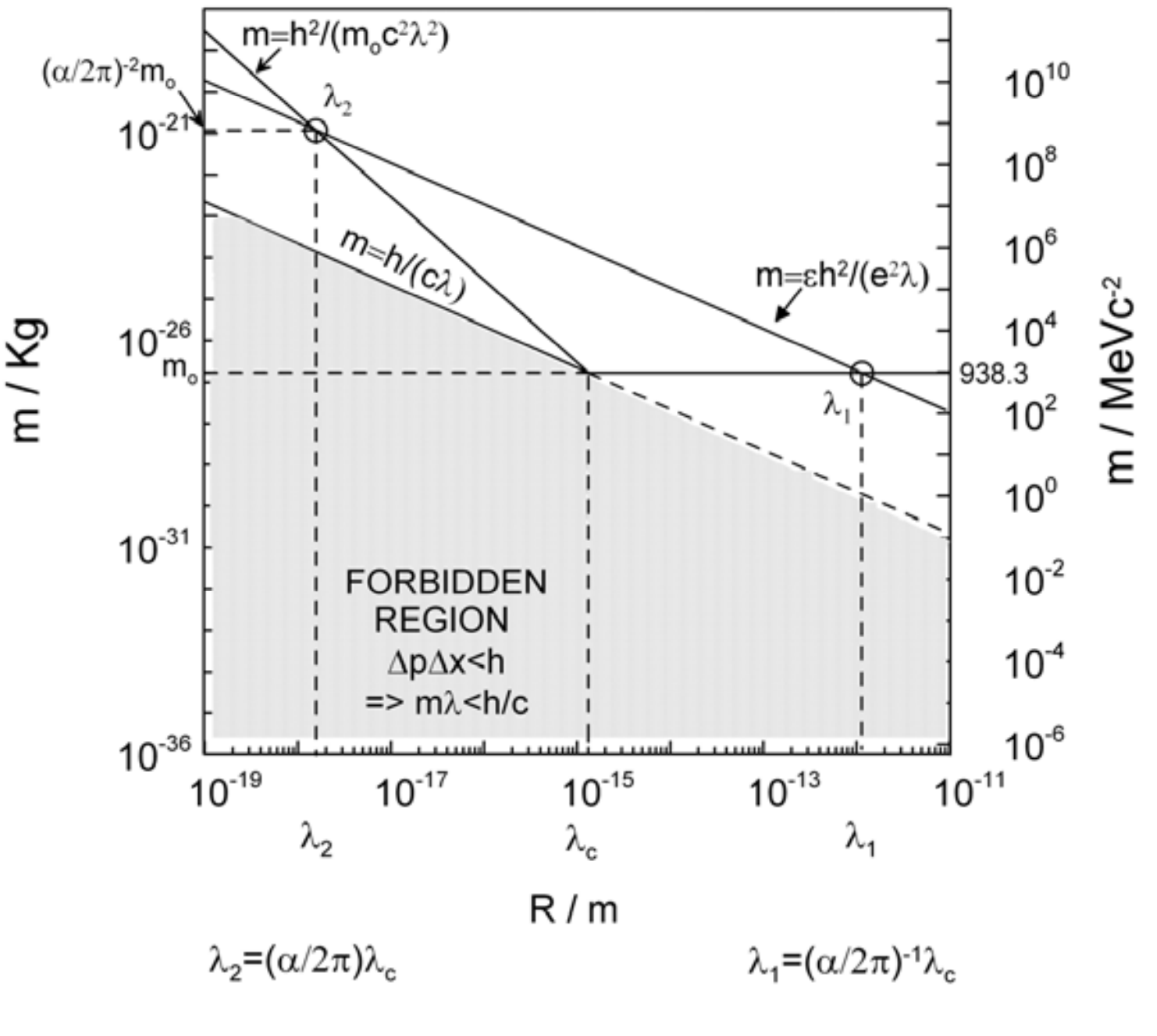

**Fig. 4.** Graphical determination of the two wavelength roots $\lambda_1$ and $\lambda_2$. They both lie on the line $m = \varepsilon h^2/e^2\lambda$ (equation (20)) obtained via combination of the de Broglie and Coulomb equations. The root $\lambda_1$ is at the intersection of this line with $m = m_o$. The root $\lambda_2$ is at the intersection with the line $m = h^2/m_o c^2\lambda$ obtained from the de Broglie equation $\lambda = h^2/(2mE_K)^{1/2}$ with $\bar{E}_K = (1/2)m_o c^2$ and also with the line $m_o = m_o(\alpha/2\pi)^{-2}$. The shaded area, obtained via $\Delta p = mc$ and $\Delta x = \lambda$, defines the region where Heisenberg's uncertainty principle is violated. The Compton length, $\lambda_c$, is the geometric mean of $\lambda_1$ and $\lambda_2$.

Setting $\eta = m/m_o$ and using equation (19) as an equality, one obtains a second root, which is the minimum $\lambda$ value satisfying equation (19):

$$\lambda_2 = \frac{e^2}{\varepsilon m_o c^2} (= (\alpha/2\pi)\lambda_c = 1.53\cdot 10^{-18}\ \text{m}) \ ; \ \bar{E}_{C,2} = -\left(\frac{3}{2}\right)\bar{E}_{attr,min} = (1/2)m_o c^2 = 469\ \text{MeV} \qquad (22)$$

with

$$\eta = \lambda_c^2/\lambda_2^2 = m/m_o = (\alpha/2\pi)^{-2} \qquad (23)$$

This root denoted $\lambda_2$ in Figure 4, can also be obtained from equation (20) by setting $m = m_o(\alpha/2\pi)^{-2}$.

Since the ratio $m/m_o$ equals[1] the Lorentz factor $\gamma\left(= (1-v^2/c^2)^{-1/2}\right)$, one can identify the factor $\eta$ with $\gamma$. In fact, equation (23) has also been derived using special relativity.[19]

The second root $\lambda_2(=1.53\cdot 10^{-18}\ \text{m})$ falls in the size range of quarks,[23,24] while the first root $\lambda_1(\sim 10^{-12}\text{-}10^{-10}\ \text{m},$ depending on the dielectric constant $\varepsilon_r)$ clearly corresponds to proton interactions in chemical and electrochemical systems,[19,25] Table 1. This phenomenon of bistability, i.e. two acceptable roots $\lambda_1$ and $\lambda_2$, is rather common in chemical kinetics and the prevailing root depends on the previous history of the system.[26,27]

**Table 1**. Proton wavelengths and energies in chemical (γ=1) and nuclear (γ=(α/2π)$^{-2}$) environments (Equations (20), (21) and (22))

| $\lambda(=\varepsilon h^2/\gamma m_o e^2)$ $\quad E_C(=\gamma m_o e^4/2\varepsilon^2 h^2)$ | | | |
|---|---|---|---|
| | $\lambda_1$ | $\bar{E}_{C,1}$ | |
| $\gamma = 1$<br>$\varepsilon_r = 1$ | $1.14\cdot 10^{-12}$ m | 635 eV | |
| $\varepsilon_r = 78$ | $0.89\cdot 10^{-10}$ m | 0.10 eV | Chemical-electrochemical systems |
| $(\lambda_1\lambda_2)^{1/2} = \lambda_c = 1.32\cdot 10^{-15}$ m | | | Compton wavelength<br>$\lambda_c = h/m_o c$ |
| | $\lambda_2$ | $\bar{E}_{C,2}$ | **Nuclear systems** |
| $\gamma = (\alpha/2\pi)^{-2}$<br>$\varepsilon_r = 1$ | $1.53\cdot 10^{-18}$ m | 469 MeV | |

Equations (20) and (21) have been recently derived and used in the treatment of proton tunnelling in chemical systems, such as polymeric proton conducting membranes.[25] These equations, which give both $\overline{E}_C$ and $\overline{E}_{attr,min}$, are very similar with the expressions for the Bohr radius and energy of the H atom. The difference is that here h rather than $\hbar$ appears due to the vibrational vs rotational motion and the sign of $\overline{E}_C$ is positive due to the repulsive proton-proton interaction.

Interestingly, it follows from (21) and (22) that:

$$(\lambda_1\lambda_2)^{1/2} = \lambda_c \tag{24}$$

, i.e. the Compton length of the proton, which is the proton size determined from scattering experiments, is the geometric mean of the two proton wavelengths $\lambda_1$ and $\lambda_2$.

*2.4. Comparison with the relativistic approach*

As already noted, when the root $\lambda_2 (< \lambda_c)$ prevails, it follows from Heisenberg's uncertainty principle that (equation (23)):

$$m = m_o\gamma = m_o(\alpha/2\pi)^{-2} = 7.41\cdot10^5\ m_o \tag{25}$$

which suggests that gravitational forces may be non-negligible in a nuclear environment, as anticipated by previous theoretical considerations accounting for relativity.[10-12]

In fact, using the definition of the Lorentz factor $\gamma = (1 - v^2/c^2)^{-1/2}$ and expressing the kinetic energy of the Hamiltonian as $(1/2)m_o v^2$ (since the $^4$He nucleus is at rest, the kinetic energy of each nucleon oscillator cannot exceed $(1/2)m_o c^2$) and the Coulombic potential energy $\overline{E}_C$ using equation (20), as was done in Ref. 19, one obtains:

$$(1/2)m_o v^2 = \frac{m_o e^4}{2\varepsilon^2 h^2}(1 - v^2/c^2)^{-1/2} \quad ; \quad \frac{v^2}{c^2}\left(1 - \frac{v^2}{c^2}\right)^{1/2} = \frac{e^4}{\varepsilon^2 c^2 h^2} = (\alpha/2\pi)^2 \tag{26}$$

which has two roots for $v^2/c^2$ between 0 and 1, i.e.

$$v^2/c^2 = (\alpha/2\pi)^2 \quad ; \quad \gamma_1 \approx 1 \quad ; \quad v = (\alpha/2\pi)c \tag{27}$$

$$v^2/c^2 = 1-(\alpha/2\pi)^4 \quad ; \quad \gamma_2 = (\alpha/2\pi)^{-2} \quad ; \quad v \approx c \tag{28}$$

where the first $\gamma$ root $(\approx 1)$, i.e. $m = m_{o,He}$, leads to the wavelength root $\lambda_1$ (Equation (20)), and the second $\gamma$ root $(=(\alpha/2\pi)^{-2})$ leads to the wavelength root $\lambda_2$ (Equation (21)). In the case of the first root, $\lambda_1$, it is $v = (\alpha/2\pi)c$, similar to $v = \alpha c$ for the case of the electron in the H atom.[23]

It is worth noting that in the present derivation, equation (23) has been obtained without using relativity, although relativity certainly provides the means to interpret physically the key result $m/m_o = (\alpha/2\pi)^{-2}$, due to the high vibrational velocity of each nucleon relative to the center of mass of the nucleus.

## 3. Velocity dependent interaction energies

The previous analysis, which has shown the existence of the root $\lambda_2$ of size $10^{-18}$ m, i.e. in the size range of color quarks,[23,24] has treated the nucleons as wave-particles vibrating with respect to the center of mass of the $^4He$ nucleus, without examining any additional vibrational motion in the interior of the nucleons which, as already shown in Figure 3, are treated as vibrating strings. Thus the previous analysis also provides an estimate for the size, $\lambda_2$, of these quark strings.

Also in the previous analysis no discussion was made about the exact nature of the attractive strong interaction energy, $\bar{E}_{attr}$, which keeps the $^4He$ nucleus together and counterbalances the Coulombic repulsion $\bar{E}_C$.

Both these items are addressed in this section. First, in close analogy[14] with equation (1), we make the hypothesis, to be proven later, that the attractive strong interaction energy, $\bar{E}_{attr}$, can be expressed in the form:

$$\bar{E}_{attr} = -\frac{Gm_{o,He}^2 f(\bar{E}_{K,s})}{R} \tag{29}$$

where G is the gravitational constant, to be determined, and $f(\bar{E}_{K,s})$ is a function of the kinetic energy, $\bar{E}_{K,s}$, of the vibrating closed string with respect to its geometric center. The kinetic energy $\bar{E}_{K,s}$ is different from the kinetic energy $\bar{E}_K$, first introduced in equation (6), as it refers to the kinetic energy of the oscillating string inside the particle $(|x| \le \lambda/2)$ (x is the distance from the geometric center of the string (Fig. 3)) with respect to the particle itself, while $\bar{E}_K = h^2/2m\lambda^2$ refers to the kinetic energy of the particle with respect to the center of mass of the He nucleus (Figure 3). Thus, also using $R = 2\lambda$, one can rewrite equation (17) as:

$$H_{osc}(\lambda, x) = \left( \frac{h^2}{2m\lambda^2}, \frac{2(e^2/\varepsilon) - 3Gm_o^2 f(\bar{E}_{K,s})}{2\lambda} \right) \tag{30}$$

As already noted, the function $f(\bar{E}_{K,s})$ is expected to express any relativistic increase in nucleon mass and gravitational interaction energy for small $|x|$ values where the velocity, v, and kinetic energy per unit string length, denoted $E'_{K,s}(x)$, becomes maximum $(E'_{K,s,max})$ (Figure 5).

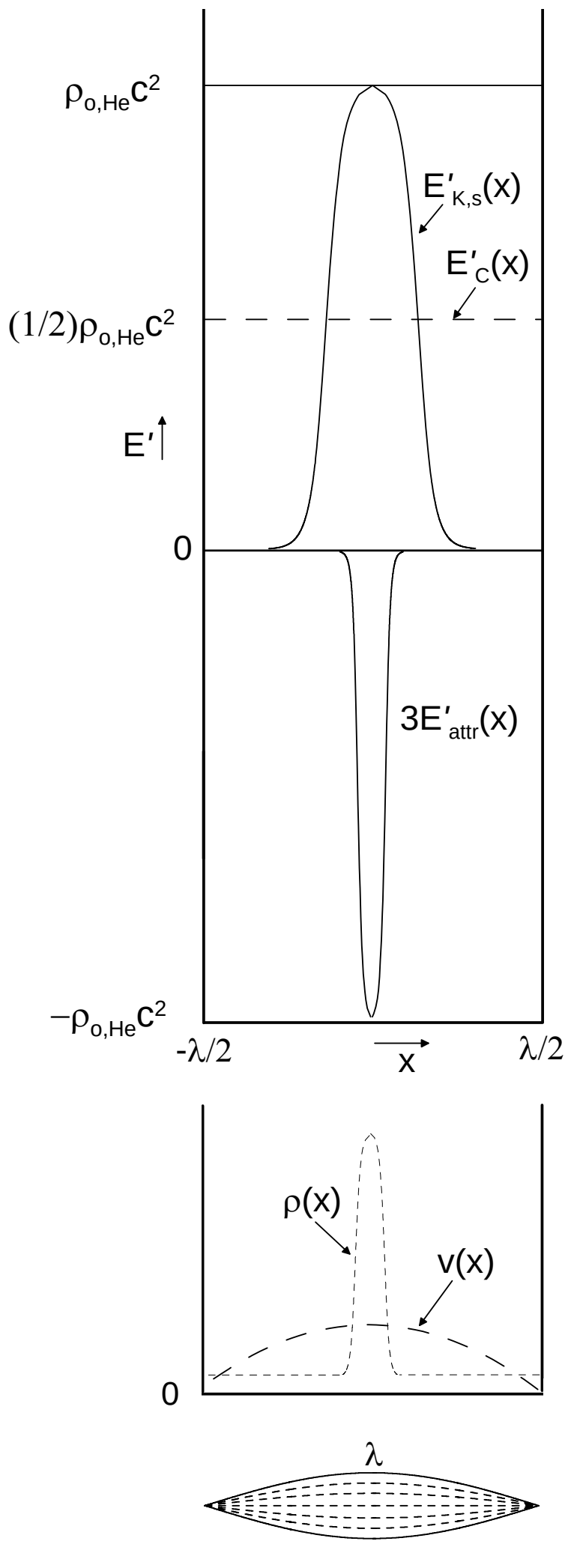

**Fig. 5.** Schematic of the dependence on distance, x, from the string center, of the kinetic energy per unit string length, $E'_{K,s}(x)$, of the total attractive interaction per unit string length, $3E'_{attr}(x)$ , of the linear string density, $\rho(x)$, and of string velocity, v(x). The figure shows the limiting case of equation (33), i.e. $E'_{K,s}(0) \approx \rho_{o,He}c^2$, thus $\rho(0) \gg \rho_{o,He}$.

We also allow for the possibility that the attractive interaction energy per unit string length, $E'_{attr}$, is also a function of x, denoted $E'_{attr}(x)$, since the magnitude of this energy will, in general, depend on the local linear mass density $\rho(x)$ and the latter may increase significantly from its rest value if $v_{max}$, and thus $E'_{K,s,max}$, reaches relativistic $(v_{max} \approx c)$ values. Thus the minimum value of $E'_{attr}(x)$, $E'_{attr}(0) = E'_{attr,min}$, corresponds to the maximum velocity and kinetic energy at $x = 0$ (Figure 5). Thus, the total kinetic energy of the string, $\bar{E}_{K,s}$, and the total attractive interaction energy, $\bar{E}_{attr}$, are given by:

$$\bar{E}_{K,s} = \int_{-\lambda/2}^{\lambda/2} E'_{K,s}(x)dx \quad ; \quad \bar{E}_{attr} = \int_{-\lambda/2}^{\lambda/2} E'_{attr}(x)dx \qquad (31)$$

The Coulombic energy per unit string length, $E'_C$, does not depend on x, thus $\bar{E}_C = \lambda E'_C$.

Thus, denoting by $E'$ the total proton or neutron energy per unit string length and utilizing equation (6) or (9) one obtains:

$$E' = \rho_{o,p,He}(x)c^2 + E'_C + 3E'_{attr}(x) + E'_{K,s}(x) = \rho_{o,He}c^2 \qquad (32)$$

where $\rho_{o,p,He}$ is the proton rest mass per unit string length. At $|x| = \lambda/2$ both $E'_{K,s}(x)$ and $3E'_{attr}(x)$ vanish, thus, in analogy with equation (8), it is $\rho_{o,p,He}c^2 = \rho_{o,He}c^2 - E'_C$. At $x = 0$, using $E'_{K,s,max} \le \rho_{o,He}c^2$ and $-3E'_{attr,min} = 2E'_C = E_{K,s,max}$ as in equation (12), one has:

$$-3E'_{attr,min} = 2E'_C = E_{K,s,max} \le \rho_{o,He}c^2 \qquad (33)$$

again with $\rho_{o,p,He}c^2 = \rho_{o,He}c^2 - E'_C$. Figure 5 corresponds to the limiting case $E'_{K,s,max} = \rho_{o,He}c^2$ . Analytical expressions for the functions $E'_{K,s}(x)$ and $E'_{attr}(x)$ for $\lambda = \lambda_2$ are presented in sections 4 and 5.

## *4. Quark strings*

The physical picture emerging from the previous analysis is shown in Figure 6. We started by searching for the size, $\lambda$, of the proton and neutron wavepackets, or strings, which in view of the starting equations (16) to (18) satisfy equation (19).

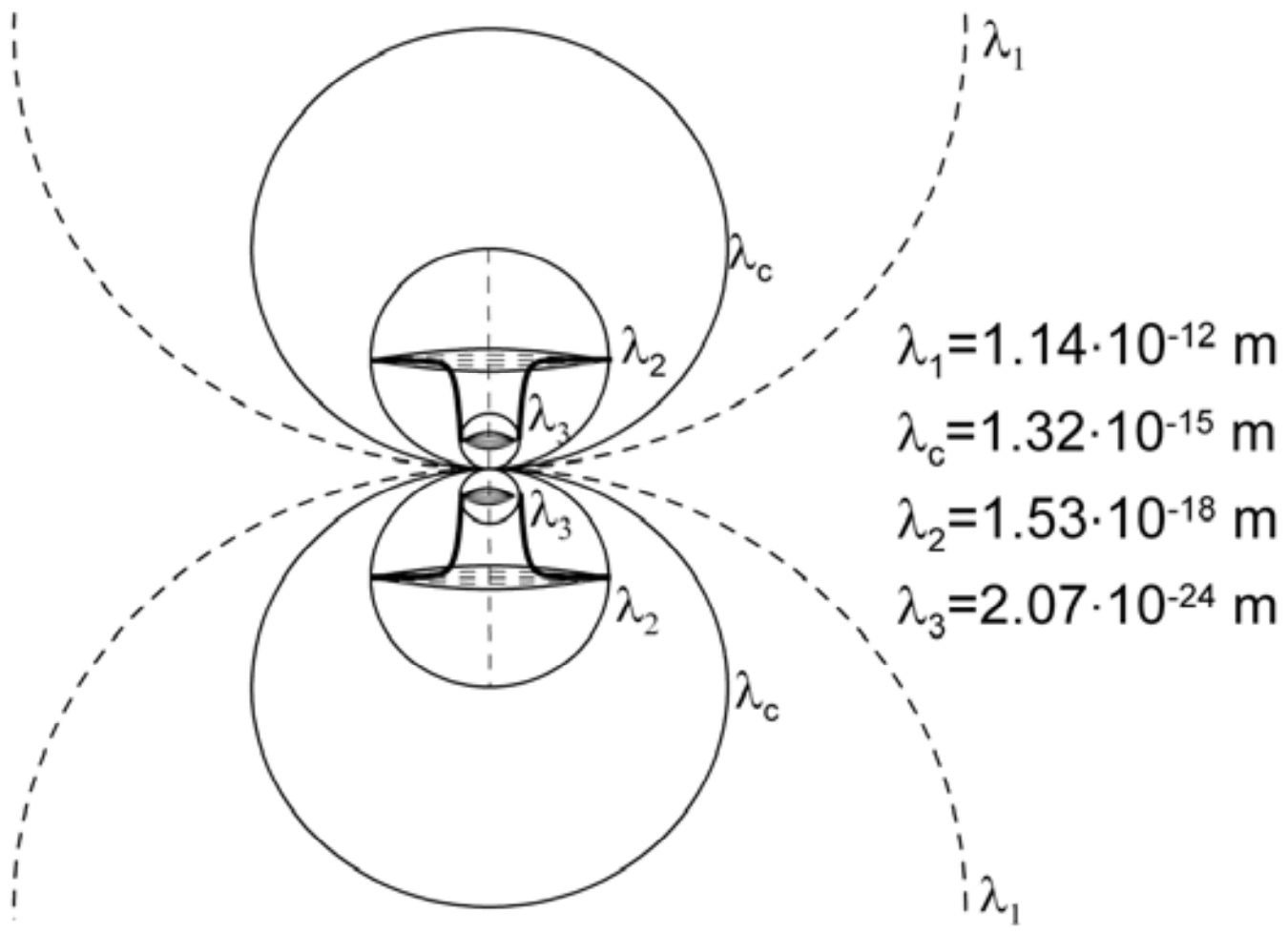

**Fig. 6.** Schematic of the interaction between two baryons, showing the wavelengths $\lambda_1(=(\alpha/2\pi)^{-1}\lambda_c)$, $\lambda_c$, $\lambda_2(=(\alpha/2\pi)\lambda_c)$ and $\lambda_3(=(\alpha/2\pi)^3\lambda_c)$ and corresponding strings of lengths $\lambda_2$ and $\lambda_3$. The gravitational interaction between the two $\lambda_3$ strings is similar to a gluon interaction, with an energy $-(1/5)(1/3)m_{o,He}c^2 = (2/15)(e^2/\varepsilon\lambda_2)$ (Equations (40) and (51)).

We found two roots $\lambda_1$ and $\lambda_2$, above and below the Compton wavelength $\lambda_c$ which satisfy equation (19). The former, $\lambda_1$, clearly refers to protons in chemistry ($\lambda_1 \sim 10^{-10} - 10^{-12}$ m, $v = (\alpha/2\pi)c$, $\gamma = 1$, Table 1). The second root, ($\lambda_2 \sim 10^{-18}$ m), corresponds to the size range of quarks[23,24] and to a total mass $\gamma m_o = (\alpha/2\pi)^{-2} m_o$. Thus for an outside observer the rest (internal oscillation free) mass of the vibrating waveparticle string is $\gamma m_o = (\alpha/2\pi)^{-2} m_o$ (Figure 6). But this quark waveparticle consists of an oscillating string with a kinetic energy per unit length, given by $E'_{K,s}(x) = E'_{K,max}\cos^2\left(\frac{\pi x}{\lambda}\right)$ and a corresponding Lorentz factor $\gamma(x)$. Because $\gamma$ deviates from unity only for v values very close to the maximum string velocity $v_{max}$, we consider a small segment $\Delta x = \delta\lambda_2$ of the vibrating string at its midpoint, such that, due to $v(x) \approx v_{max}$, thus $\rho(x) \gg \rho_{o,He}$, it contains practically all the mass $(\alpha/2\pi)^{-2} m_o$ of the vibrating string. We will return to this assumption and make the necessary corrections in the next section.

Thus, one can rewrite equations (19), but this time with a rest mass $(\alpha/2\pi)^{-2}m_o$:

$$\frac{h^2}{2m_o(\alpha/2\pi)^{-2}\lambda^2}=\frac{e^2}{2\varepsilon\lambda}\le\frac{1}{2}m_o(\alpha/2\pi)^{-2}c^2 \qquad (34)$$

Similarly to equation (19), this equation has two roots. These roots are shown in Figure 7. The first root coincides with $\lambda_2$, i.e.

$$\lambda_2=\frac{e^2}{\varepsilon m_o c^2}\ (=(\alpha/2\pi)\lambda_c) \quad ; \quad \overline{E}_{C,2}=\frac{m_o(\alpha/2\pi)^{-2}e^4}{2\varepsilon^2h^2}=\frac{1}{2}(\alpha/2\pi)^{-2}m_oc^2 \qquad (35)$$

Similarly the second root, denoted $\lambda_3$, is given by:

$$\lambda_3=\frac{e^2}{\varepsilon m_o(\alpha/2\pi)^{-2}c^2} \quad (=(\alpha/2\pi)^3\lambda_c=(\alpha/2\pi)^4\lambda_1)$$
$$\overline{E}_{C,3}=\frac{m_o(\alpha/2\pi)^{-4}e^2}{2\varepsilon^2h^2}=\frac{1}{2}(\alpha/2\pi)^{-4}m_oc^2 \qquad (36)$$

Thus the oscillating string of the quark wavepacket with size $\lambda_2$ corresponds to a wavepacket, or oscillating string, of size $\lambda_3=\delta\lambda_2=(\alpha/2\pi)^3\lambda_c=2.07\cdot10^{-24}$ m, rest mass $(\alpha/2\pi)^{-2}m_o$ and relativistic mass $(\alpha/2\pi)^{-4}m_o$ (Figures 6, 7 and 8).

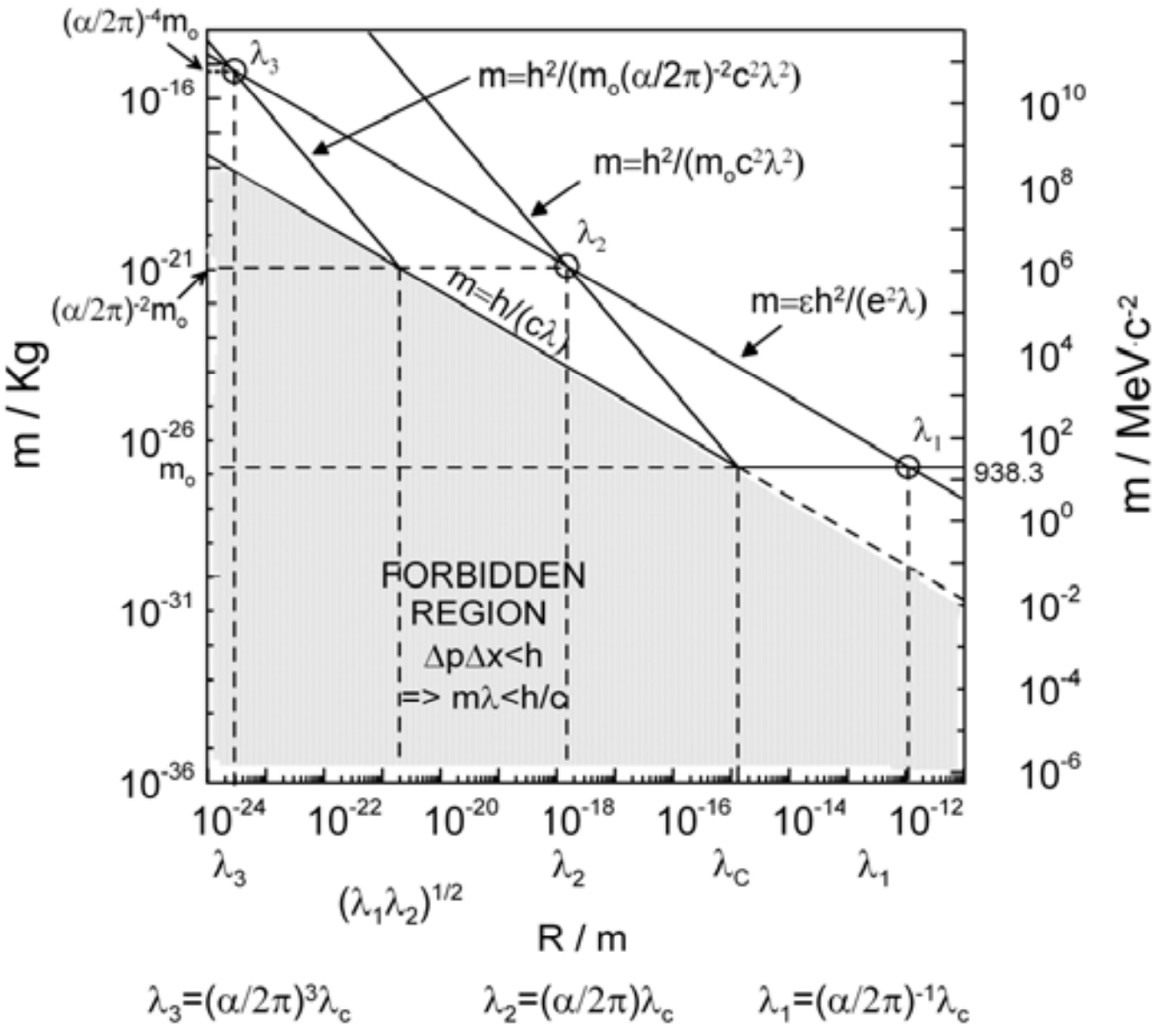

**Fig. 7.** Graphical determination of the wavelength root $\lambda_3$, which lies at the intersection of the de Broglie-Coulomb $m=eh^2/e^2\lambda$ line (Figure 4) and the line $m=h^2/(m_o(\alpha/2\pi)^{-2}c^2\lambda^2)$ obtained from the de Broglie equation $\lambda=h^2/(2m\overline{E}_K)^{1/2}$ with $\overline{E}_K=(1/2)m_o(\alpha/2\pi)^{-2}c^2$. It also lies on the $m=m_o(\alpha/2\pi)^{-4}$ line.

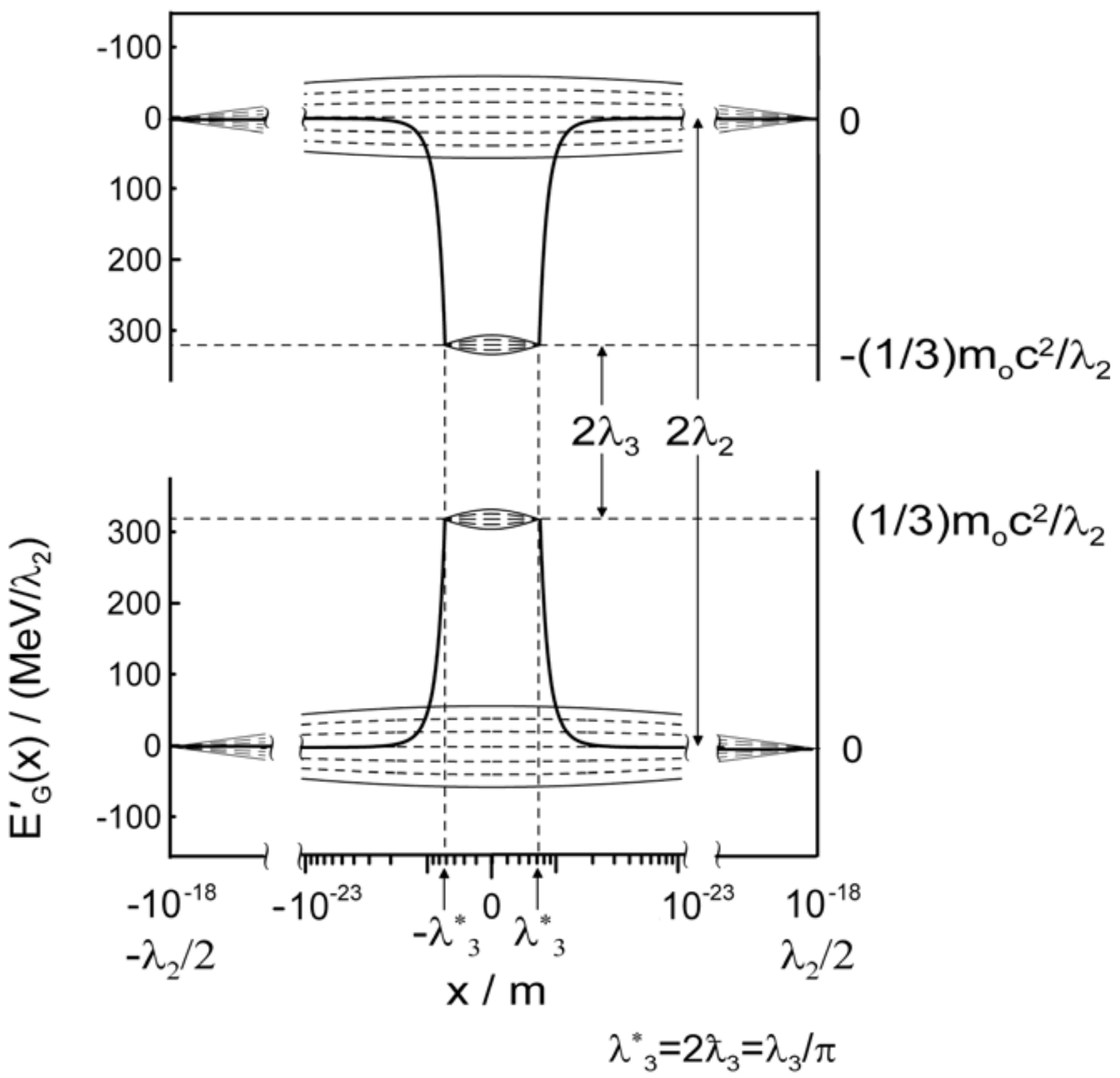

**Fig. 8.** Dependence on x of the gravitational interaction energy per unit string length, $E'_G(x)$, between two quark strings of length $\lambda_2$ (from equations (44) and (47)). The relativistic increase in linear density for $v \approx c$ at small $|x|$ values creates a string of size $\lambda_3^* = 2\bar{\lambda}_3 \approx \lambda_3$ containing most of the string mass. The gravitational interaction energy $\overline{E}_G \left(= \lambda_2 \int_{-\lambda_2/2}^{\lambda_2/2} E'_G(x)dx\right)$ between the two quark string equals $(1/15)m_o c^2$. The interaction bears all the features of a gluon interaction.

Figures 6 and 8, the latter depicting the actual gravitational energy per unit string length, $E'_G(x)$ vs x dependence obtained in the next section, provide a picture very similar to that envisioned for a gluon interaction.[23,24] The two interacting strings of size $\lambda_3$, lying inside the protons and neutrons (size $\lambda_c$) and inside the quarks (size $\lambda_2$) approach each other, (Figure 6 and 8) due to their gravitational attraction to distances $2\lambda_3$ where the gravitational attraction is of the size of strong nuclear interactions, as shown next.

***5. Gravitational constant***

*5.1. A first estimate of G*

We denote by $G_{max}$ the value of the gravitational constant to be determined under the previously made assumption that all the mass of the string of length $\lambda_2$ is concentrated at $x = 0$ and we consider the gravitational force, $F_{G,max}$, exerted between two string waveparticles at a distance $2\lambda_2$ (Figure 8):

$$-F_{G,max} = \frac{G_{max} m^2}{4\lambda_3^2} = \frac{G_{max} m_o^2 (\alpha/2\pi)^{-8}}{4\lambda_2^2 (\alpha/2\pi)^4} = \frac{G_{max} m_o^2}{4\lambda_2^2}(\alpha/2\pi)^{-12} = \\ = \frac{G_{max} m_o^2}{4\lambda_2^2}\gamma_{max}^6 = \frac{G_{max} m_o^2}{4\lambda_2^2}(1.66\cdot 10^{35}) \qquad (37)$$

where we denote $\gamma_{max} = \left(1 - \frac{v_{max}^2}{c^2}\right) = (\alpha/2\pi)^{-2}$ and $v_{max}$ is the maximum string vibrational velocity at $x = 0$. Thus the gravitational force exerted between the two string waveparticles is a factor $(\alpha/2\pi)^{-12} (= 1.66\cdot 10^{35})$ larger than the Newtonian force between two masses $m_o$ at a distance $\lambda_2$, in absence of relativistic effects, i.e.:

$$-F_{G,o} = \frac{G m_o^2}{4\lambda_2^2} \quad ; \quad \frac{F_{G,max}}{F_{G,o}} = (G_{max}/G)(\alpha/2\pi)^{-12} \qquad (38)$$

$$E_{G,max}/E_{G,o} = F_{G,max}/F_{G,o} = (G_{max}/G)(\alpha/2\pi)^{-12} \qquad (39)$$

i.e. $E_{G,max}$ is a factor $1.6\cdot 10^{35}$ larger than that anticipated from Newton's Law.

Recalling equation (15), i.e., $\bar{E}_{attr} = -(2/3)\bar{E}_C = (1/3)m_{o,He}c^2$, utilizing equation (22), i.e. $\bar{E}_C = e^2/2\varepsilon\lambda_2$, and identifying $E_{G,max}$ as $\bar{E}_{attr}$, one thus obtains:

$$-E_{G,max} = \frac{G_{max} m_o^2 (\alpha/2\pi)^{-12}}{2\lambda_2} = \frac{(2/3)(e^2/\varepsilon)}{2\lambda_2} = (1/3)m_{o,He}c^2 \qquad (40)$$

which gives:

$$G_{max} = \left(\frac{2}{3}\right)(\alpha/2\pi)^{12}\frac{(e^2/\varepsilon)}{m_o^2} \qquad (41)$$

Substituting the values of the constants one obtains:

$$G_{max} = 3.3371 \cdot 10^{-10}\ m^3 kg^{-1} s^{-2} \tag{42}$$

which happens to be exactly a factor of five larger than the experimental G value[21,22] $(6.6742 \cdot 10^{-11}\ m^3 kg^{-1} s^{-2})$. This is quite reasonable in view of the fact that $G_{max}$ corresponds to the assumed hypothetical situation where the entire string mass is concentrated at the midpoint $(|x| \leq \lambda_3)$ of the string.

*5.2 Exact computation*

It follows from equations (37) to (39) that $-E_{G,max} = \dfrac{G_{max} m_o^2}{2\lambda_1} \gamma_{max}^6$ and since $E'_{G,max}\lambda_2 = \overline{E}_{G,max}$, it follows:

$$-E'_{G,max} = \frac{G_{max} m_o^2}{2\lambda_1 \lambda_2} \gamma_{max}^6 \tag{43}$$

where $\gamma_{max} = \left(1 - \dfrac{v_{max}^2}{c^2}\right)^{-1/2}$ and $v_{max}$ is the maximum string velocity, corresponding to $|x| \approx \lambda_3 = (\alpha / 2\pi)^2 \lambda_2 \approx 0$. For $|x|$ values in the interval $(\lambda_3, \lambda_2)$, the corresponding $E'_G(x)$ value is given by:

$$-E'_G(x) = \frac{G_{max} m_o^2}{2\lambda_1 \lambda_2} \gamma^6 = \frac{G_{max} m_o^2}{2\lambda_1 \lambda_2} \left(1 - \frac{v^2(x)}{c^2}\right)^{-3} \tag{44}$$

The velocity profile of the string is given by:

$$\frac{v^2}{v_{max}^2} = \cos^2\left(\frac{\pi x}{\lambda_2}\right) \tag{45}$$

Thus from (43), (44) and (45)

$$\frac{E'_G(x)}{E'_{G,max}} = \frac{\gamma_{(x)}^6}{\gamma_{max}^6} = \left(\frac{\left(1 - \dfrac{v_{(x)}^2}{c^2}\right)}{\left(1 - \dfrac{v_{max}^2}{c^2}\right)}\right)^{-3} = \frac{1}{\left[1 + (\gamma_{max}^2 - 1)\left(1 - \cos^2(\dfrac{\pi x}{\lambda_2})\right)\right]^3} \tag{46}$$

and thus, accounting for $\gamma_{max}^2 \gg 1$ and $\pi x/\lambda_2 \ll 1$, thus $\cos^2(\pi x/2\lambda_2) \approx 1 - \left(\frac{\pi x}{\lambda_2}\right)^2$ it is:

$$\frac{E'_G(x)}{E'_{G,max}} = \left(\frac{\lambda_2}{\pi x}\right)^6 \gamma_{max}^{-6} = \left(\frac{\lambda_3^*}{x}\right)^6 \quad ; \quad \lambda_3^* = \frac{\lambda_2}{\pi\gamma_{max}} = \frac{\lambda_3}{\pi} = 2\bar{\lambda}_3 \tag{47}$$

Thus $E'_G(x)$ equals $E'_{G,max}$ at $x = 2\bar{\lambda}_3 = 2(\alpha/2\pi)^2\bar{\lambda}_2 \approx 0$. A plot of the function $E'_G(x)$ accounting for $E'_{G,max} = -E_{G,max}/\lambda_2 = (1/3)m_{o,He}c^2$ (Equation (40)) is given in Figure 8. Therefore the actual gravitational energy of interaction, $\bar{E}_G$, is given by:

$$\bar{E}_G = \int_{\bar{\lambda}_3^*}^{\lambda_2} E'_G(x)\,dx = E'_{G,max}\int_{\bar{\lambda}_3^*}^{\lambda_2}\left(\frac{\lambda_3^*}{x}\right)^6 dx = \frac{1}{5}(\bar{\lambda}_3^*)E'_{G,max} = \frac{1}{5}E_{G,max} \tag{48}$$

where we have accounted for $\lambda_2^{-5} \ll (\lambda_3^*)^{-5}$. Thus $\bar{E}_G$ equals one fifth of the value it would have if all the string mass was confined in $|x| \le \lambda_3$.

Thus,

$$\frac{\bar{E}_G}{E_{G,max}} = \frac{G}{G_{max}} = \frac{1}{5} \tag{49}$$

and combining with equation (40) one obtains:

$$Gm_o^2 = \left(\frac{1}{5}\right)\left(\frac{2}{3}\right)(\alpha/2\pi)^{12}\left(\frac{e^2}{\varepsilon}\right) \tag{50}$$

thus:

$$G = \left(\frac{2}{15}\right)\left(\frac{e^2}{\varepsilon m_o^2}\right)(\alpha/2\pi)^{12} = \left(\frac{2}{15}\right)\left(\frac{e^2}{\varepsilon m_o^2}\right)\left(\frac{e^2}{\varepsilon ch}\right)^{12} = 6.6742\cdot 10^{-11}\ \mathrm{m^3kg^{-1}s^{-2}} \tag{51}$$

Upon substituting ($e = 1.602\cdot 10^{-19}$ C, $\varepsilon = 1.11\cdot 10^{-10}\ \mathrm{C^2/Nm^2}$, $m_o = 1.666157\cdot 10^{-27}$ kg $= 934.643$ MeV/c$^2$, $h = 6.6236\cdot 10^{-34}$ J$\cdot$m, $c = 2.997\cdot 10^8$ m/s, one obtains $G = 6.6742\cdot 10^{-11}\ \mathrm{m^3kg^{-1}s^{-2}}$, which is in excellent quantitative agreement with the currently CODATA recommended value[22] of $6.6742 \pm 0.001 \times 10^{-11}$ $\mathrm{m^3kg^{-1}s^{-2}}$ (Figure 9).

The above $m_o$ value is the mean between the average of the proton and neutron rest masses $(1.67377 \cdot 10^{-27}\ \mathrm{kg})$ and the average mass of a nucleon in the $^{56}Fe$ or $^{98}Mo$ (or $^{40}Ca$) nuclei. The former two are the elements most abundant in the attracting masses used in the experimental measurement of the gravitational constant.[21,22] We note that, as is evident from equation (51), there is a minor effect (in the fourth significant figure) of the chemical composition of the attracting masses on the computed or measured value of the gravitational constant. This is known experimentally,[21,22] and is one reason for the significant scattering of the experimental data shown in Figure 9.

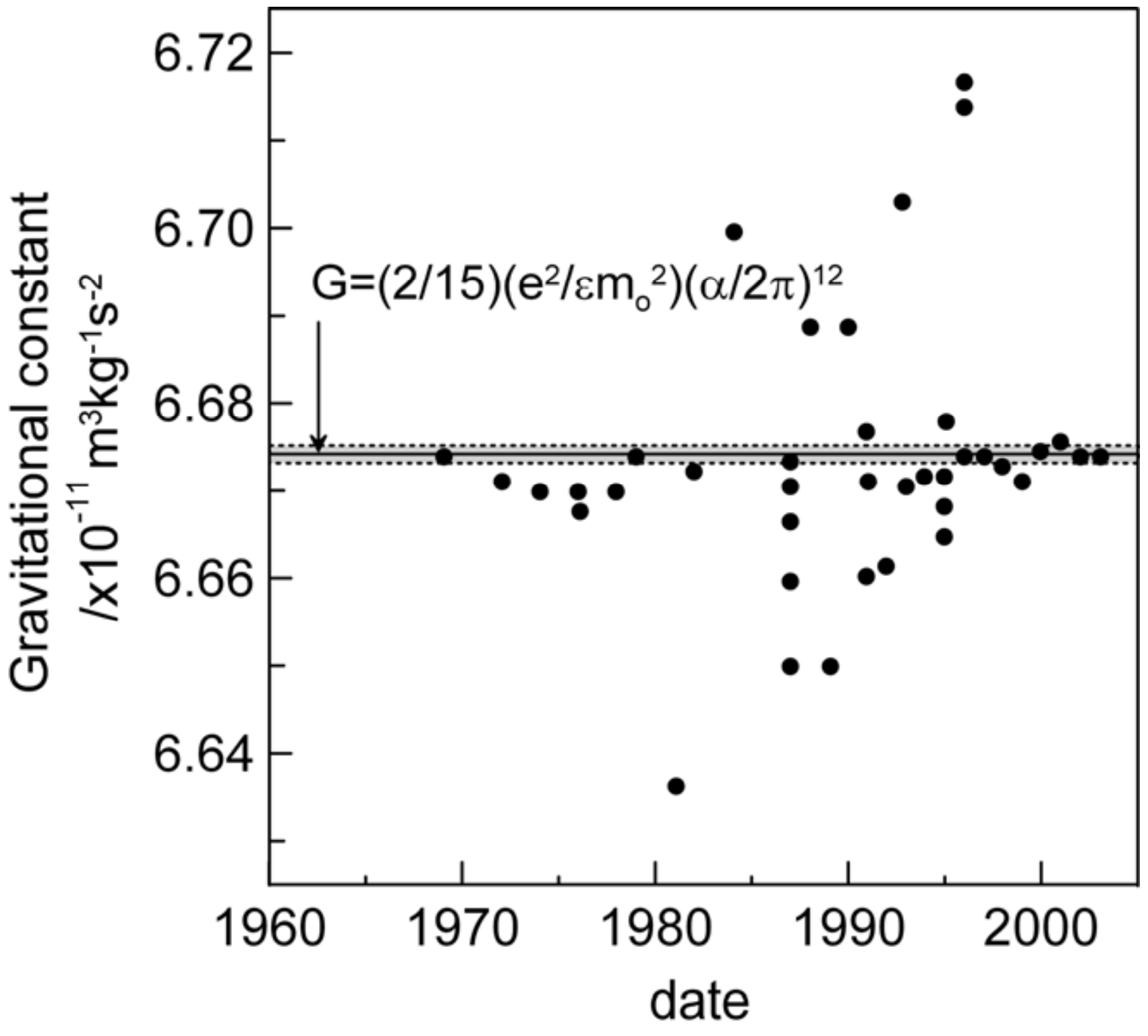

**Fig. 9.** Comparison of equation (51) with the time evolution of the experimental gravitational constant value and with some of the most recent experimental values;[21,22] Pre-1997 values data from Table 2 in Ref. 21. Post-1997 values from Table X in Ref. 22. Shaded areas shows the CODATA recommended[22] value of $(6.6742 \pm 0.001) \times 10^{-11}$ $\mathrm{m^3kg^{-1}s^{-2}}$.

Using the computed gravitational constant G value one can then compare the nuclear gravitational energy of interaction of two protons, i.e. $Gm_o^2(\alpha/2\pi)^{-12}/R$ with the energy of the electrostatic repulsion, $e^2/\varepsilon R$, and with the energy, $Gm_o^2/R$ of the gravitational interaction outside the nuclear environment (Figure 10). One observes that the nuclear gravitational energy is 2/15 of the repulsive Coulombic energy, as is evident from equation (51), and that the transition from Newton's Law to the relativistic Newton Law occurs between $\lambda_2$ and $\lambda_3^* \approx \lambda_3$, according to equation (44) or (47). The relativistic Newton's Law for $R \le \lambda_2$ may be associated with the strong interaction forces, as previously discussed.

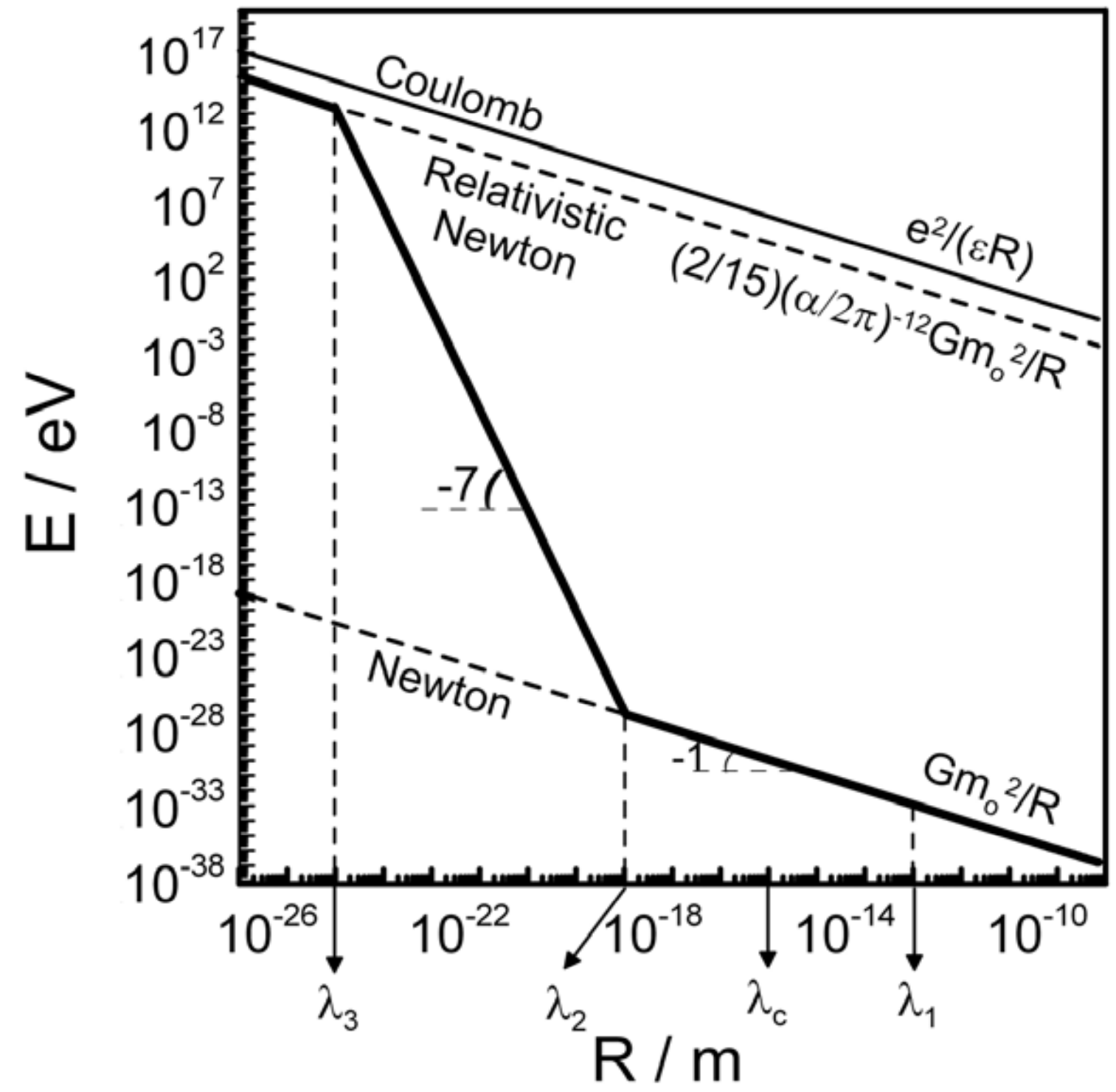

**Fig. 10.** Distance dependence of the Coulombic and gravitational energy between two protons. The thick line shows the gravitational energy and depicts the transition from Newton's law $-Gm_o^2/R$ to the relativistic Newton (or strong interaction) line $-(2/15)(\alpha/2\pi)^{-12}Gm_o^2/R$ (equation (51)). The transition occurs between $\lambda_2$ and $\lambda_3^*(\approx \lambda_3)$ as described by equations (44) or (47). The relativistic Planck length, $R_{Pl,rel}$, is at $4.27\,\lambda_3$ (eq. (54)).

### *5.3. Planck length*

The convergence of electromagnetic, gravitational and strong interaction forces, shown in Figure 10 to occur at a distance $\sim\lambda_3(= 2.07\cdot 10^{-24}\ \text{m})$ is commonly thought to take place at the Planck length, $R_{Pl}(=(\hbar G/c^3)^{1/2} \approx 1.62 \cdot 10^{-35}\ \text{m})$. Here we show that $\lambda_3$ can indeed be viewed as a relativistic Planck length.

First we note that, using the definition of the Planck mass, $m_{Pl}(=(\hbar c/G^3)^{1/2} = 2.18 \cdot 10^{-8}\ \text{kg})$ and equation (51), one can rewrite equation (51) in the form:

$$G = \hbar c / m_{Pl}^2 \quad ; \quad m_{Pl} = m_o\left((15/2)\alpha^{-1}(\alpha/2\pi)^{-12}\right)^{1/2} = m_o(15/2\alpha)^{1/2}(\alpha/2\pi)^{-6} \quad (52)$$

Second we note that when examining equation (51) in conjunction with $\gamma = (\alpha/2\pi)^{-2}$ and $m = m_o\gamma$, as was done in Ref. 19, i.e. without accounting for the quark string and concomitant approach of the relativistic masses at a distance $2\lambda_3$, then one concludes that the gravitational constant in the nucleus departs from its macroscopic value G to a relativistic value, $G_{rel}$, given by:

$$G_{rel} = G\gamma^4 = G(\alpha/2\pi)^{-8} \quad (53)$$

In fact equation (53) can be justified by invoking the relativistic effects on length and time,[19] and this is an equivalent way to view equation (51). Upon introducing equation (53) in the definition of the Planck length one obtains:

$$R_{Pl,rel} = \left(\frac{\hbar G_{rel}}{c^3}\right)^{1/2} = \left(\frac{2}{15\alpha}\right)^{1/2}\lambda_3 = 4.27\lambda_3 \quad (54)$$

which shows that indeed $\lambda_3$ can be viewed as a relativistic Planck length. Interestingly, by combining equations (53), (54) and $\lambda_3 = (\alpha/2\pi)^3\lambda_c$ one obtains

$$R_{Pl} = \left(\frac{2}{15\alpha}\right)^{1/2}(\alpha/2\pi)^7\lambda_c = 1.62 \cdot 10^{-35}\ \text{m} \quad (55)$$

Using the definition of the proton or neutron Compton length, $\lambda_c = h/m_o c$ in equation (55) one obtains:

$$m_o = \left(\frac{2}{15\alpha}\right)^{1/2}(\alpha/2\pi)^7\frac{h}{R_{Pl}c} \quad (56)$$

which defines the proton mass in terms of the Planck length in a way similar to the second equation (52), which defines $m_o$ in terms of the Planck mass $m_{Pl}$.Thus, given the values of the physical constants h, c, e and G, then the rest proton or neutron mass, $m_o$, is defined. But this is also obvious from equation (51).

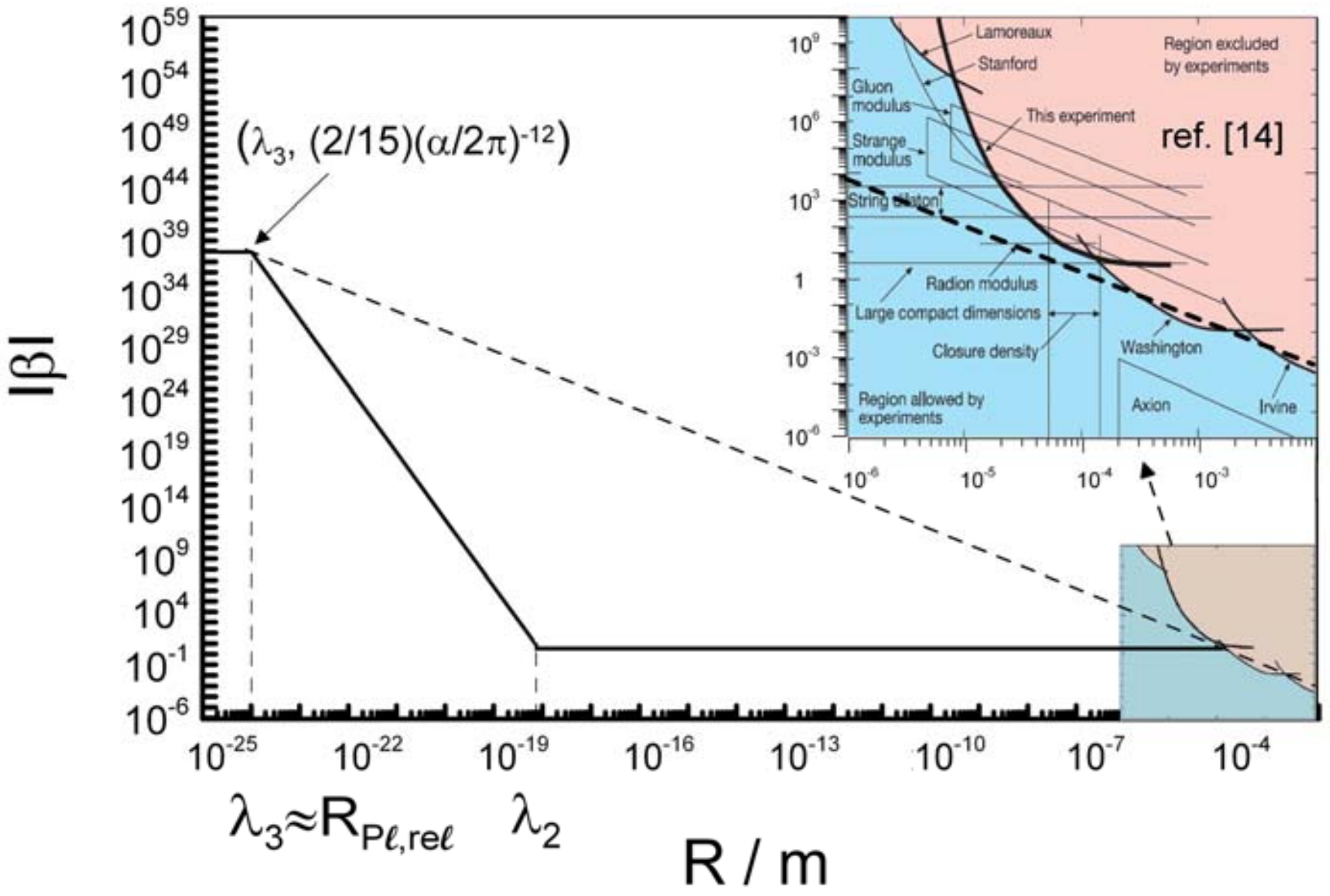

**Fig. 11.** Comparison of the computed strength of modulus force $\beta$ of equation (1) i.e., $\beta = (2/15)(\alpha/2\pi)^{-12} = 2.22 \cdot 10^{34} (= f_{max})$ at $\lambda_3 \approx R_{Pl,rel}$ with the upper limits (inset) set on $\beta$ by previous studies.[14] Inset obtained from Ref. 14.

It is interesting to note that the computed via equation (51) magnitude of the gravitational moduli force, β, of equation (1) which corresponds to the relativistic Planck length $R_{Pl,rel}$ of equation (54) lies close to the extension of the line established as an upper limit by previous theoretical and experimental studies[14] as shown in Figure 11.

## *6. Nuclear binding energies*

### *6.1. Ion-induced dipole energies*

The previous analysis shows that the Coulombic, strong interaction and gravitational nuclear interaction energies are all of the order of magnitude, $e^2/\varepsilon\lambda_2 \approx m_o c^2 \approx 0.94\text{ GeV}$, when examined at distances below the size of quarks $(\lambda_2 = (\alpha/2\pi)\lambda_c \approx 10^{-18}\text{ m})$. This is consistent with the fact that the existence of quarks was first detected via inelastic electron scattering of protons and neutrons with electron beams of energies 1-10 GeV.[23,24] The same interaction energies, when examined at distances of the size of protons and neutrons, i.e. at $\lambda_c$, are of the order $e^2/\varepsilon\lambda_c = \alpha m_o c^2 \approx 6.9\text{ MeV}$ which is exactly the average binding energy of nucleons in nuclei.[23,24]

It thus becomes interesting to examine to what extent one can make more exact estimates of the binding energy of small nuclei (e.g. $^4$He , D) by considering the values of the above interaction energies at the Compton length of the nucleus.

For particles of size $\lambda_c$, the ion-induced dipole forces between protons and neutrons cannot be neglected,[28,29] although, interestingly, due to the concomitant increase in mass and decrease in electrical polarizability, these forces become negligible in relation to the Coulombic and other nuclear forces for particles of size $\lambda_2$ or smaller, (Appendix A).

Ion-induced dipole (ID) forces obey a $R^{-5}$ law,[28-30] i.e.:

$$E_{ID} = -\frac{1}{2}\left(\frac{e^2}{\varepsilon}\right)\frac{a_n}{R^4} \tag{57}$$

where $a_n$ is the electrical polarizability of the neutron (practically equal with that of the proton, [28,29]) which can be computed from the general expression[30]:

$$a_n^{(o)} = \frac{(e^2/\varepsilon)}{m\omega^2} = \alpha\bar{\lambda}_c^3 = 6.54\cdot 10^{-50}\ \text{m}^3 \tag{58}$$

which is exactly valid when the field is coaxial with the induced dipole.[30] Using this expression and accounting for the relativistic increase in mass one computes that at $\lambda_3$, the polarizabilities of neutron and protons are of the order $\alpha(\alpha/2\pi)^8\bar{\lambda}_c^3$ (Appendix A) and thus the ID forces are negligible in comparison with the Coulombic and gravitational forces, despite their $R^{-5}$ (vs $R^{-2}$) dependence. A comparison between the $a_n^{(o)}$ value computed from equation (58) and literature values[28,29] is also given in Appendix B.

*6.2. $^4$He nucleus*

Thus considering a $^4$He nucleus (one Coulombic, six gravitational and four ion-induced dipole interactions) one obtains the following expression for the total potential energy, $\bar{E}_T$, of the four particles at an average nucleon distance $R$.

$$\bar{E}_T(R) = \bar{E}_C + \bar{E}_{ID} + \bar{E}_G = \frac{e^2}{\varepsilon R} - \frac{e^2}{\varepsilon}\cdot\frac{2a_n}{R^4} - 6\frac{Gm^2}{R} \tag{59}$$

One can use equation (59) to plot the function $\bar{E}_T(R)$ in the interval $[ƛ_{c,He},\infty[$ (Figure 12) ($Gm_o^2 = \left(\frac{2}{15}\right)(e^2/\varepsilon)(\alpha/2\pi)^{12}$ (Equation (51)) and also using $a_n = (4/3)a_n^{(o)}$ to account for the angle between the field generated by each proton and the dipole induced in each neutron by the two protons, (Appendix C). One observes that the minimum $\bar{E}_T$ value is at $ƛ_{c,He}$ and equals -28.43 MeV, thus $E_b = -\bar{E}_T$ practically coincides with the experimental,[21,22] $\Delta mc^2 (= 28.30 \text{ MeV})$, value for the binding energy of $^4$He. Thus setting $R = ƛ_{c,He}$ in equation (59) one obtains:

$$E_{b,{}^4He} = -\bar{E}_T(ƛ_{c,He}) = -\frac{e^2}{\varepsilon ƛ_{c,He}}\left[\left(1-\frac{4}{5}\right)-\frac{2a_n}{ƛ_{c,He}^3}\right] \tag{60}$$

and using, by definition, $ƛ_{c,He} = ƛ_c/4$, and noting that $e^2/\varepsilon ƛ_{c,He} = 4\alpha m_{o,He}c^2$ one obtains:

$$E_{b,{}^4He} = 4\alpha m_{o,He}c^2\left[(1/5)-\left(\frac{4}{3}\right)(128\alpha)\right] = 28.43 \text{ MeV} \tag{61}$$

where[21,22] $m_{o,He} = 1.66747 \cdot 10^{-27}$ kg or 931.844 MeV/c$^2$.

The computed value coincides within 0.5% with the experimental $\Delta mc^2$ value of 28.30 MeV.[21,22] The value of the activation energy (1.38 MeV) computed via differentiation of equation (59) and shown in Figure 12 is also in good qualitative agreement with experiment.[23,24,31] A comparison of the function $\bar{E}_T(R)$ with the Yukawa potential is presented in Figure 12b. In the latter case the depth of the energy well has been adjusted for the case of four nucleons. One observes several similarities, but $\bar{E}_T(R)$ falls more sharply at $ƛ_{c,He}$ than the Yukawa potential.

### *6.3. Binding energy of $^2H$*

In this case it is $ƛ_{c,D} = ƛ_c/2$ and the $\bar{E}_T(R)$ function is given by:

$$\bar{E}_T(R) = -\frac{G_o m_o^2(\alpha/2\pi)^{-12}}{R} - (e^2/\varepsilon)\frac{a_n}{2R^4} \tag{62}$$

therefore, setting $a_n = a_n^{(o)} = \alpha ƛ_c^3$, one obtains:

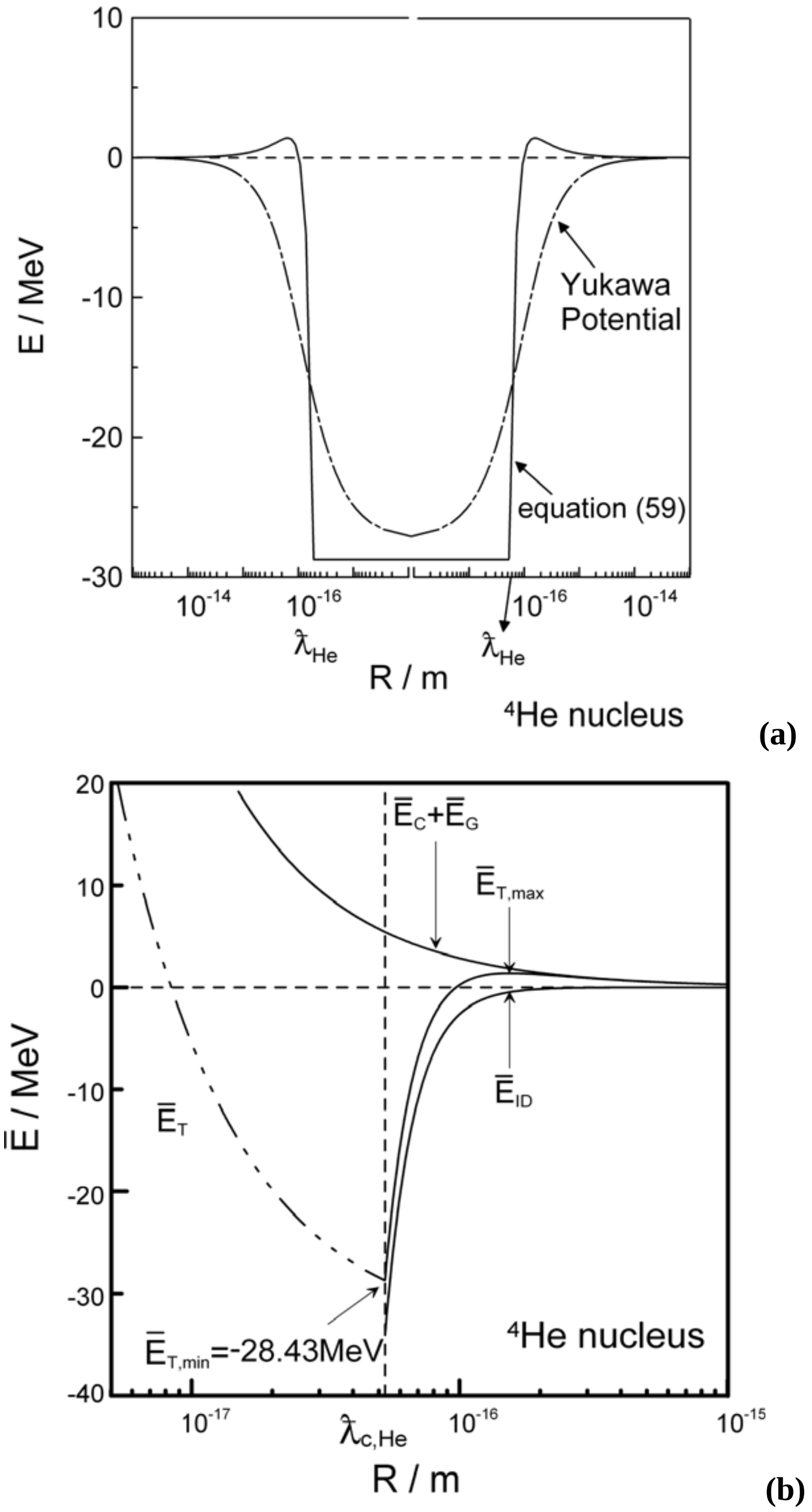

**Fig.12.** **(a)** Plot of equation (59) showing the dependence on distance, R, of the total potential energy $\overline{E}_T(R)$ of two protons and two neutrons forming a $^4He$ nucleus and comparison with the normalized (for four nucleons) Yukawa potential. **(b)** Plot of equation (59), showing the individual contributions of $\overline{E}_C + \overline{E}_G (= (1/5)E_C)$ and $\overline{E}_{ID}$ on the total potential energy $\overline{E}_T$. For $R < \lambda_{c,He}$ the ion-induced dipole interaction energy $\overline{E}_{ID}$ is maintained at its $\lambda\!\!\!^{-}_{c,He}$ value.

$$E_{b,D} = -\bar{E}_T(\lambda\!\!\!^{-}_{C,D}) = \frac{e^2}{\varepsilon \lambda\!\!\!^{-}_{C,D}}\left[-\frac{2}{15} - 4\alpha\right] = 2\alpha m_{o,D}c^2\left[\left(\frac{2}{15}\right) + 4\alpha\right] = 2.2245 \text{ MeV}$$

(63)

The computed value, $E_{b,D}$, using[21,22] $m_{o,D} = 1.67278 \cdot 10^{-27}$ kg $= 937.806$ MeV/c$^2$, is in quantitative agreement (better than 0.05%) with the experimental value[21,22] of 2.2245 MeV of the binding energy of D.

*6.4. ID energies and color quarks*

From equation (63) it follows that the ID energy in the $^2$H nucleus accounts for 18% of the total binding energy, i.e. for 0.40 MeV. It is interesting to compute the charge, q, and corresponding dipole moment, $\mu$, of the dipole. The energy, $E^o_{ID}$, corresponding to a dipole of charge e and length $\lambda\!\!\!^{-}_c$ is given by:

$$E^o_{ID} = \frac{1}{2}\frac{e^2}{\varepsilon \lambda\!\!\!^{-}_c} = \frac{1}{2}\alpha m_o c^2 = 3.415 \text{ MeV} \quad (64)$$

The equivalent mass is $3.415$ MeV/c$^2$, which practically coincides with the mass of u quarks.[23,24] In the D case it is:

$$E_{ID} = \frac{1}{2}\frac{q^2}{\varepsilon \lambda\!\!\!^{-}_c} = 0.40 \text{ MeV} \quad (65)$$

And thus from (64) and (65) it follows $(q/e)^2 = 0.117$, thus $q/e \approx \pm 0.34 \approx \pm 1/3$. The negative root gives the electric charge of d quarks and the positive root gives half the charge of u quarks. These observations suggest that u and d quarks, which are responsible in the standard model for the attractive forces between protons and neutrons, may be also related to the dipole energy and equivalent mass of the proton-neutron ion-induced dipole interactions in nuclei. At the same time it appears possible that some other types of quarks, such as the top (t) quark with a charge $+2/3$ and a very large rest mass of $170.9$ GeV$/c^2$ is related to gravitational moduli forces, e.g. to a particle of Compton length $\lambda_2$, thus of rest mass $m_o(\alpha/2\pi)^{-1}$, thus gravitational interaction energy $(1/5)m_o(\alpha/2\pi)^{-1} \approx 162$ GeV.

## 7. CONCLUSIONS

Due to Heisenberg's uncertainty principle and relativity, the high vibrational energies in nuclei cause an increase in the masses of nucleons and quarks and thus make gravitational forces quite significant. The latter appear to be closely related with the strong interaction forces between quarks. These forces counterbalance the Coulombic repulsions in nuclei at distances of the order of the relativistic Planck length $(\sim 10^{-24}\ \mathrm{m})$ and this permits the analytical computation of the gravitational constant. The present analysis, which confirms the proposition that gravitational moduli forces can be very significant,[7-14,32,33] does not treat particles as pointlike entities, but considers them as strings, i.e. as one-dimensional extended objects.[34-36] The general picture emerging from the analysis, and summarized in Figure 6, is in good qualitative agreement with the standard model, regarding the concepts of quarks, quark confinement, and gluon interactions.

At distances of the order of the Compton length ($\sim 10^{-15}$ m), ion-induced dipole interactions, which also bear several similarities with color quark interactions, also assist to counterbalance the strong Coulombic repulsions between protons and to create significant (~7 MeV/nucleon) binding energies. This allows for the computation of the binding energies of small nuclei, such as the $^{4}$He and D nuclei, in close agreement to experiment.

**Acknowledgements**

CGV acknowledges helpful discussions with Professor Ilan Riess from the Physics Department of the Technion. Sincere thanks are also expressed to Michael Tsampas for technical assistance and helpful discussions.

## APPENDIX A

First we derive equation (58) for a particle of mass $m_o$ and Compton length $\lambda_c = h/m_o c$. It is:

$$a_n^{(o)} = \frac{(e^2/\varepsilon)}{m_o\omega^2} = \frac{\frac{e^2}{\varepsilon c\hbar}\hbar c}{m_o\frac{c^2}{\bar{\lambda}_c^2}} = \alpha\bar{\lambda}_c^3 \qquad \text{(A1)}$$

Thus at $\bar{\lambda}_{c,He}(=\bar{\lambda}_c/4)$ the ratio of ID to Coulombic energies in a $^4$He nucleus is:

$$-\frac{E_{ID}}{E_C} = \frac{4\left(\frac{1}{2}\right)\left(\frac{e^2}{\varepsilon}\right)\frac{\alpha\bar{\lambda}_c^3}{4^4\bar{\lambda}_c^4}}{\frac{e^2}{4^{-1}\varepsilon\bar{\lambda}_c}} = 128\alpha \approx 1 \qquad \text{(A2)}$$

Then we apply equation (58) for a particle of rest mass $m_o(\alpha/2\pi)^{-4}$ and thus a Compton length $h/(m_o(\alpha/2\pi)^{-4}c) = (\alpha/2\pi)^4\bar{\lambda}_c$. It is:

$$a_n^{(o)} = \frac{e^2/\varepsilon}{m_o(\alpha/2\pi)^{-4}\omega^2} = \frac{\frac{e^2}{\varepsilon c\hbar}\hbar c}{m_o(\alpha/2\pi)^{-2}\frac{c^2}{(\alpha/2\pi)^4\bar{\lambda}_c^2}} = \alpha(\alpha/2\pi)^8\bar{\lambda}_c^3 \qquad \text{(A3)}$$

Thus at $\bar{\lambda}_{c,He}(\alpha/2\pi)^2(=\bar{\lambda}_c(\alpha/2\pi)^2/4)$ the ratio of ID to Coulombic energies is:

$$-\frac{E_{ID}}{E_C} = \frac{4\left(\frac{1}{2}\right)\left(\frac{e^2}{\varepsilon}\right)\frac{\alpha(\alpha/2\pi)^8\bar{\lambda}_c^3}{4^{-4}\bar{\lambda}_c^4(\alpha/2\pi)^8}}{\frac{e^2}{4^{-1}\varepsilon\bar{\lambda}_c(\alpha/2\pi)^4}} = 128(\alpha/2\pi)^4 \ll 1 \qquad \text{(A4)}$$

## APPENDIX B

Proton and neutron electric polarizabilities computed from proton scattering experiments $(1.2\cdot 10^{-48} \pm 30\%\ m^3)$[28,29] are a factor of $18 \pm 30\%$ larger than the value computed from the theoretical[30] equation (58), i.e. $6.54\cdot 10^{-50}\ m^3$. This difference can be understood as follows:

When neglecting the attractive gravitational energy in the computation of the electrical polarizability of neutrons from proton scattering experiments,[28,29] one assigns to the ID interactions the entire attractive energy. Denoting by $E^*_{ID}$ this, overestimated, ID energy, and by $a^*_n (= 1.2 \cdot 10^{-48} \pm 30\%\ m^3)$ [28,29] the corresponding polarizability, it is:

$$E^*_{ID} = E_{ID} + E_G \quad ; \quad \frac{E^*_{ID}}{E_{ID}} = 1 + \frac{E_G}{E_{ID}} \tag{B1}$$

It is therefore

$$\frac{a^*_n}{a^{(o)}_n} = 1 + \frac{E_G}{E_{ID}} \tag{B2}$$

and taking into account equation (A1), i.e. $a^{(o)}_n = \alpha \lambda^3_C$, and equation (50) it follows:

$$-E_G = (2/15)\frac{e^2}{\varepsilon R} \quad ; \quad E_{ID} = \frac{a^{(o)}_n}{2R^4}\left(\frac{e^2}{\varepsilon}\right) = \frac{\alpha \lambda^3_c}{2R^4}\left(\frac{e^2}{\varepsilon}\right) \tag{B3}$$

Therefore $$\frac{a^*_n}{a^{(o)}_n} = \left(\frac{4}{15}\right)\alpha^{-1}\frac{R^3}{\lambda^3_C} \tag{B4}$$

which for $R = \lambda_c$ gives $a^*_n / a^{(o)}_n$=37, which is factor of two larger than the experimental $a^*_n / a^{(o)}_n$ ratio. Agreement is exact for $R / \lambda_c = 0.8$.

## APPENDIX C

***Neutron polarizability in the $^4$He nucleus:***

The energy stored per induced dipole is given by:

$$(1/2)q^2 / \varepsilon d = E_{ID} \tag{C1}$$

When the field is coaxial with the induced dipole, then $E_{ID}$ can also be expressed via:

$$E_{ID} = (1/2)a^{(o)}_n \varepsilon |E|^2 = (1/2)a^{(o)}_n \frac{e^2}{\varepsilon R^4} \tag{C2}$$

where $|E|$ is the modulus of the field $\vec{E}$.

When the field forms an angle $\theta$ with the induced dipole, then

$$E_{ID} = (1/2)\vec{\mu} \cdot \vec{E} = (1/2)a_n \varepsilon |E|^2 \cos^2\theta = (1/2)a_n \frac{e^2}{\varepsilon R^4} \qquad \text{(C3)}$$

where $\vec{\mu}$ is the dipole moment $q\vec{d}$.

From (C2) and (C3) (since $E_{ID}$ is the same, and given by equation (C1) in both cases) it follows $a_n = a_n^{(o)} / \cos^2\theta$. For the case of the $^4$He nucleus, it is $\theta \approx 30°$, for a tetrahedral arrangement, thus $\cos^2\theta = 3/4$, thus $a_n = (4/3)a_n^{(o)}$.

## APPENDIX D

**List of Symbols and constant values**

| | |
|---|---|
| $a_n$ | electrical polarizability, $m^3$ |
| $a_n^{(o)}$ | $a_n$ computed from equation (53), $m^3$ |
| c | speed of light, $2.997925\times10^{8}$ m/s |
| e | unit charge, $1.6021765\times10^{19}$ C |
| E | energy, J or MeV |
| $\bar{E}$ | time-averaged kinetic or interaction energy, J or MeV |
| $E'$ | energy per unit string length, J/m or MeV/m |
| $\bar{E}'$ | time-averaged value of energy per unit length, J/m or MeV/m |
| G | gravitational constant, $6.6742\times10^{-11}$ $m^3kg^{-1}s^{-2}$ |
| $G_{max}$ | G estimate defined in equation (37), $m^3kg^{-1}s^{-2}$ |
| h | Planck constant, $6.6260693\times10^{-34}$ Js |
| $\hbar$ | $h/2\pi$ |
| H | Hamiltonian, J or MeV |
| $\langle H \rangle$ | time averaged value of Hamiltonian, J or MeV |
| $m_{Pl}$ | Planck mass, $(hc/G^3)^{1/2}$, kg |
| $m_{p,o}$ | $=1.67262171\times10^{-27}$ kg |
| $m_{n,o}$ | $=1.67492728\times10^{-27}$ kg |
| $m_{o,w}$ | $=1.6591543\times10^{-27}$ kg |
| $m_o$ | $=1.666157\times10^{-27}$ kg |
| $m_{o,He}$ | $=1.66747\times10^{-27}$ kg |
| R | interparticle distance, m |
| $R_{Pl}$ | Planck length, $(hG/c^3)^{1/2}$, m |
| x | distance from string center, m |
| V | combined potential energy of Newtonian gravity and modulus force, defined in equation (1), J or MeV |
| v | velocity |

**Greek symbols**

| | |
|---|---|
| $\alpha$ | fine structure constant, $e^2/\varepsilon c\hbar$, 1/137.035=7.297353×10$^{-3}$ |
| $\alpha/2\pi$ | 1.161410×10$^{-3}$ |
| $(\alpha/2\pi)^{12}$ | 6.02324×10$^{-36}$ |
| $\beta$ | parameter expressing the magnitude of the modulus force, defined in equation (1) |
| $\gamma$ | time-averaged Lorentz factor $(1-v^2/c^2)^{-1/2}$ |
| $\varepsilon$ | $4\pi\varepsilon_o\varepsilon_r = 1.112649\times10^{-10}\varepsilon_r\ C^2/Nm^2$ |
| $\varepsilon_o$ | permittivity of vacuum, $8.854187817\times10^{-12}\ C^2/Nm^2$ |
| $\varepsilon_r$ | relative dielectric constant |
| $\eta$ | $m/m_o$ ratio $(=\gamma)$ defined in equation (23) |
| $\lambda$ | wavelength, m |
| $\lambda_c$ | Compton length, $h/m_oc$ |
| $\bar{\lambda}$ | $\lambda/2\pi$ |
| $\pi$ | 3.1415926 |
| $\rho_1$, $\rho_2$ | volume mass densities, equation (1) |
| $\rho$ | linear mass density of a string, kg/m |
| $\rho_o$ | linear rest mass density of a string, kg/m |

**Subscripts**

| | |
|---|---|
| attr | attractive interaction energy |
| b | binding energy |
| c | Compton (for Compton length) |
| C | Coulombic interaction energy |
| K | kinetic energy |
| n | neutron |
| o | rest mass or linear density |
| osc | oscillator energy or Hamiltonian |
| p | proton |
| P | potential energy |
| rel | relativistic |
| s | string |

**Superscripts**

| | |
|---|---|
| — | time-averaged energy |
| ′ | quantity per unit string length |